\documentclass[a4paper,fleqn,usenatbib]{mnras}

\usepackage[T1]{fontenc}
\usepackage{ae,aecompl}

\usepackage{subfigure}
\usepackage{graphicx}	
\usepackage{amsmath}	
\usepackage{amssymb}	
\usepackage{color}
\usepackage{mathptmx}
\usepackage{epsfig}
\usepackage{wasysym}
\usepackage{xfrac}
\usepackage{pifont}
\usepackage[table,xcdraw]{xcolor}
\usepackage{booktabs}
\usepackage{multirow}
\usepackage{comment}
\usepackage{soul}
\usepackage{tikz}
\usetikzlibrary{positioning,decorations.pathreplacing,shapes}

\newcommand\basez{0.5}

\title[The assembly of dwarf galaxies]{The role of mergers and interactions in driving the evolution of dwarf galaxies over cosmic time}

\author[G. Martin et al.]{G. Martin,$^{1,2}$\thanks{E-mail: garrethmartin@arizona.edu}, R. A. Jackson,$^{3}$ S. Kaviraj,$^{3}$ H. Choi,$^{4}$ J. E. G. Devriendt,$^{5}$ Y. Dubois,$^{6}$ \newauthor  T. Kimm,$^{4}$  K. Kraljic,$^{7}$ S. Peirani,$^{6,8}$ C. Pichon,$^{6,9}$ M. Volonteri,$^{6}$ and S. K. Yi$^{4}$
\\
$^{1}$Steward Observatory, University of Arizona, 933 N. Cherry Ave, Tucson, AZ 85719, USA.\\
$^{2}$Korea Astronomy and Space Science Institute, 776 Daedeokdae-ro, Yuseong-gu, Daejeon 34055, Korea.\\
$^{3}$Centre for Astrophysics Research, School of Physics, Astronomy and Mathematics, University of Hertfordshire, Hatfield, AL10 9AB, UK. \\
$^{4}$ Department of Astronomy and Yonsei University Observatory, Yonsei University, 50 Yonsei-ro, Seodaemun-gu, Seoul 03722, Republic of Korea. \\
$^{5}$ Dept of Physics, University of Oxford, Keble Road, Oxford OX1 3RH, UK.\\
$^{6}$ Institut d'Astrophysique de Paris, CNRS \& UPMC, UMR 7095, 98 bis Boulevard Arago, F-75014 Paris, France. \\
$^{7}$ Institute for Astronomy, University of Edinburgh, Royal Observatory, Edinburgh, EH9 3HJ, UK.\\
$^{8}$ Universit\'e C\^ote d'Azur, Observatoire de la C\^ote d'Azur, CNRS, Laboratoire Lagrange, Nice, France.\\
 $^{9}$ Korea Institute for Advanced Study (KIAS) 85 Hoegiro, Dongdaemun-gu, Seoul, 02455, Republic of Korea.
}

\pubyear{2020}
\begin{document}
\label{firstpage}
\pagerange{\pageref{firstpage}--\pageref{lastpage}}
\maketitle

\begin{abstract}
    Dwarf galaxies ($M_{\star}<10^{9}$~M$_{\odot}$) are key drivers of mass assembly in high mass galaxies, but relatively little is understood about the assembly of dwarf galaxies themselves. Using the \textsc{NewHorizon} cosmological simulation ($\sim40$~pc spatial resolution), we investigate how mergers and fly-bys drive the mass assembly and structural evolution of around 1000 field and group dwarfs up to $z=\basez$. We find that, while dwarf galaxies often exhibit disturbed morphologies (5 and 20 per cent are disturbed at $z=1$ and  $z=3$ respectively), only a small proportion of the morphological disturbances seen in dwarf galaxies are driven by mergers at any redshift (for $10^{9}$~M$_{\odot}$, mergers drive under 20 per cent morphological disturbances). They are instead primarily the result of interactions that do not end in a merger (e.g. fly-bys). Given the large fraction of apparently morphologically disturbed dwarf galaxies which are not, in fact, merging, this finding is particularly important to future studies identifying dwarf mergers and post-mergers morphologically at intermediate and high redshifts. Dwarfs typically undergo one major and one minor merger between $z=5$ and $z=0.5$, accounting for 10 per cent of their total stellar mass. Mergers can also drive moderate star formation enhancements at lower redshifts (3 or 4 times at $z=1$), but this accounts for only a few per cent of stellar mass in the dwarf regime given their infrequency. Non-merger interactions drive significantly smaller star formation enhancements (around two times), but their preponderance relative to mergers means they account for around 10 per cent of stellar mass formed in the dwarf regime.
\end{abstract}




\section{Introduction}


Dwarf galaxies ($M_{\star}<10^{9}$~M$_{\odot}$) are the most abundant systems in the Universe, regardless of redshift, \citep[e.g.][]{Fontana2006,Karachentsev2013,Grazian2015} and are key drivers of the mass assembly and evolution of massive haloes \citep{Press1974}. While there is a large body of observational and theoretical work on the assembly of high mass galaxies \citep[e.g.][]{Zepf1989,Hopkins2010,Dokkum2010,Ferreras2014,Robotham2014,Rodriguez2015,Company2016,Man2016,Martin2018_progenitors,Padmanabhan2020}, relatively little is understood about the role of interactions and mergers in the evolution and assembly of dwarfs, despite the fact that the hierarchical assembly of galaxies, which is a key prediction of $\Lambda$CDM \citep[e.g.][]{Fall1980,Bosch2002,Agertz2011}, is also thought to extend to dwarf mass haloes \citep[e.g.][]{Wheeler2015}.

Until recently, observational \citep[e.g.][]{Bradac2009,Graham2012,Martiznez2012,Huang2016,Janz2017,Richtler2018} and theoretical \citep[e.g.][]{DOnghia2008,Sawala2010,Cloet2014,GarrisonKimmel2019} studies of dwarf galaxy assembly and structural evolution have remained quite limited in scope, in particular due to the observational challenges involved in detecting dwarf galaxies outside of the local volume. The SDSS, which has provided much of the discovery space for astronomers in the local and intermediate redshift Universe, is incomplete beyond an $r$-band effective surface-brightness of $\sim$23~mag~arcsec$^{-2}$ \citep[e.g.][]{Driver2005,Blanton2005,Zhong2008,Bakos2012,Williams2016} since the majority of these objects are much fainter than this limit \citep{Haberzettl2007, Martin2019} bespoke reductions \citep[e.g.][]{Stierwalt2015,Fliri2016} or purpose built instruments \citep[e.g.][]{Abraham2014} are required to study them. There also exist significant theoretical challenges in simulating large enough volumes to provide a realistic cosmological context while simultaneously resolving galaxies down to the dwarf regime. Until recently, high resolution simulations of dwarf mass galaxies have been limited to cosmological zoom-in simulations which simulate a small number of isolated haloes at high resolution embedded within a low resolution environment \citep[e.g.][]{Governato2010,Wang2015,Onorbe2015}. Together, these observational and theoretical barriers have limited our ability to understand the evolution and properties of dwarf galaxies in a statistical sense.

Despite these challenges, a body of observational evidence is beginning to emerge regarding the importance of mergers and interactions for the assembly and evolution of dwarf galaxies. While dwarf galaxies are routinely observed as the satellites of more massive galaxies in the local Universe \citep[e.g.][]{McConnachie2012,Sales2013,Gaia2018,Carlsten2020}, there is also compelling observational evidence that halo substructures persist down to the lowest masses. Indirect evidence exists in the form of disrupted morphologies or other signatures of previous interactions \citep{Rich2012, Martiznez2012,Johnson2013} and observations of dwarf galaxy pairs and mergers within more massive haloes \citep[e.g.][]{Tully2006,Ibata2013,Crnojevic2014,Deason2014b,Paudel2017}. Such observations are relatively scant, however, owing to the limited depths of contemporary surveys \citep[e.g.][]{Driver2005,Blanton2005}. There also exists direct observational evidence from isolated merging dwarf galaxies or gravitationally bound isolated clusters of dwarf galaxies \citep{Sales2013,Stierwalt2017,Privon2017,Besla2018}, which represent a powerful validation to the hierarchical paradigm of galaxy assembly in the low mass regime. Additionally, we are also beginning to understand the impact of mergers and interactions on the structure and assembly histories of dwarfs by studying, for example, the star formation rates and gas distributions of isolated dwarf pairs \citep[e.g.][]{Noeske2001,Stierwalt2015,Pearson2016,Privon2017,Pearson2018} and appealing to such processes also has the potential to resolve apparent conflicts between recent observations and the $\Lambda$CDM paradigm in the dwarf regime \citep[e.g.][]{Shin2020,Jackson2020b,Montes2020}.

Now, large improvements in the design, sensitivity and field of view of modern instruments are beginning to alleviate some of the challenges present in previous studies. Next generation instruments, like the James Webb Space Telescope, the Dragonfly Telephoto Array the Hyper Suprime-Cam and the Legacy Survey of Space and Time (LSST) from the Vera Rubin Observatory \citep[][]{Olivier2008,Merritt2014,Robertson2017,Aihara2018} will enable new detections of dwarf galaxies. In particular detailed observations of dwarf galaxies at high redshift will soon become possible. While studies of lensed systems in the HST frontier fields \citep{Bradac2009,Atek2015,Huang2016,Castellano2016,Atek2018} are already capable of probing delensed stellar masses as low as $10^{7}$~M$_{\odot}$, the JWST NIRCam imager will be capable of 10 to 100 times the sensitivity of HSC with angular resolution in the infrared better than 0.1 arcsec, corresponding to sub kpc resolutions at all redshifts \citep{Beichman2012}. The work of \citet{Patej2015} suggest that it may be possible to directly detect between 20 per cent to 70 per cent of Local Group dwarf analogues out to $z=7$ and other theoretical work \citep[e.g.][]{Williams2018,2018Cowley,Behroozi2020} suggests that JWST will detect thousands of $M_{\star}<10^{7}$ galaxies even beyond $z\sim10$ with many more detected at low and intermediate redshifts. At lower redshifts, instruments like the Vera Rubin Observatory will push the detection limits of contemporary wide area surveys to unprecedented depths (5 sigma depth up to 27.5 mags in the $r$-band).

At the same time, advances in computing power and numerical techniques have enabled a new generation of high resolution zoom-in and cosmological simulations. While a majority of new cosmological simulations \citep[][]{Tremmel2017,Pillepich2018,Dave2019,Nelson2019} still have insufficient spatial resolution ($>100$~pc), zoom-in simulations, which simulate a handful of objects at very high (10 - 50~pc) resolution \citep[e.g.][]{Wang2015,Hopkins2018,GarrisonKimmel2019b} are now capable of adequately resolving the physics of low mass dwarf galaxies. Very high resolution cosmological simulations which simulate controlled volumes with a realistic cosmological context \citep[e.g.][]{Dubois2020} are now allowing for robust comparison with theory in the dwarf regime.

With these new methods, an understanding of the assembly of dwarf galaxies across a significant fraction of cosmic time may be within reach. However, great care will be required in the interpretation of this new data, especially since common assumptions used in studies of high mass galaxies may not hold true in the dwarf regime. Evidence of the assembly history of galaxies are encoded in their shapes and can be used to infer a galaxy's recent interaction history, particularly through the detection of visible morphological disturbances and asymmetries in galaxies \citep{Conselice2003}. It is important, however, to understand how the role of the different mechanisms that produce these asymmetries evolve with galaxy mass and redshift. For example, since a wide range of different mechanisms can drive either temporary or permanent transformations in the morphologies of galaxies, including major mergers \citep[e.g.][]{Toomre1977,Negroponte1983,DiMatteo2007,Hopkins2009,Ferreras2009,Conselice2009,Taranu2013,Naab2014,Deeley2017}, minor mergers \citep[e.g.][]{Dekel2009,Kaviraj2014b,Fiacconi2015,Zolotov2015,Welker2017,martin2018_sph,Jackson2020}, fly-bys and tidal stripping \citep{Miller1986, Moore1998, Abadi1999,Manodeep2012,Kim2014,LangL2014,Choi2017} or, particularly in the early Universe, internal processes \citep[e.g.][]{Bournaud2008,Agertz2009,Forster2011,Cibinel2015,Hoyos2016}. Understanding these processes is likely to be especially important at high redshifts, where objects will be too poorly resolved to easily distinguish the processes that produce asymmetries and morphological disturbances in their light profile. 

In this paper we make use of the \textsc{NewHorizon} cosmological hydrodynamical simulation \citep{Dubois2020}, which models a 10~Mpc radius spherical volume with stellar mass and maximum spatial resolutions of 10$^4$ M$_{\odot}$ and 34~pc, typically seen in some state-of-the-art zoom-in simulations of individual haloes. We attempt to understand how major and minor mergers as well as non-merger interactions like fly-bys drive morphological changes in galaxies as a function of their stellar mass and redshift. We also make predictions for how these processes drive the mass assembly of dwarf galaxies through direct accretion of stellar mass formed ex-situ and through the enhancement star formation. The structure of this paper is as follows: 
\begin{enumerate}
    \item In Section \ref{sec:method} we present an overview of the \textsc{NewHorizon} simulation, relevant physics and measurements, including the treatment of baryonic physics, merger trees, calculation of galaxy photometry and shapes and describe our method for identifying morphologically disturbed dwarfs.
    \item In Section \ref{sec:freq_disturbances} we investigate the drivers of morphological disturbances as a function of stellar mass. We make predictions for the morphological merger fraction and disturbed fraction (i.e. including galaxies with disturbed morphologies not due to mergers) as a function of mass and redshift as well as make predictions for merger durations in the dwarf regime.
    \item In Section \ref{sec:starbursts}, we investigate the effect of mergers and interactions on the instantaneous star formation rates of dwarf galaxies.
    \item In Section \ref{sec:assembly} we make predictions for the fraction of dwarf galaxies masses that are assembled as a result of accreted stellar mass formed ex-situ and as a result of enhancements in star formation induced by mergers and interactions.
    \item In Section \ref{sec:summary} we summarise our results.
\end{enumerate}

\section{Method}

\label{sec:method}

\subsection{The \textsc{NewHorizon} simulation}

\label{sec:NH}

The \textsc{NewHorizon} simulation\footnote{\href{http://new.horizon-simulation.org}{http://new.horizon-simulation.org}} is a zoom-in of the 142~Mpc length box of the Horizon-AGN simulation~\citep{Dubois2014}. Initial conditions are generated using cosmological parameters that are compatible with \emph{WMAP7} $\Lambda$CDM cosmology \citep{Komatsu2011} ($\Omega_{\rm m}=0.272$, $F\Omega_\Lambda=0.728$, $\sigma_8=0.81$, $\Omega_{\rm b}=0.045$, $H_0=70.4 \ \rm km\,s^{-1}\, Mpc^{-1}$, and $n_s=0.967$). Within the original Horizon-AGN volume, a spherical volume with a $10 \ \rm~Mpc$ radius and an effective resolution of $4096^3$ is defined, corresponding to a dark matter (DM) mass resolution and \textit{initial} gas mass resolution of $m_{DM}=1.2\times 10^6 \ \rm M_\odot$ and $m_{gas}=2\times 10^5 \ \rm M_\odot$. This region is chosen to have average density and probes field and group environments. The high-resolution volume is embedded within regions of increasingly coarse resolution, with DM mass resolution eventually decreasing from $10^7\,\rm M_\odot$ to $5\times 10^9\ \rm M_\odot$. Galaxies at the boundary of the high resolution volume, are removed based on the presence of low-resolution DM particles in the host halo and are not included in our analysis.

The current redshift reached by the simulation is $z=0.25$. In this study, we use a base redshift of $z=0.5$ and consider the evolution of objects from $z=6$ up to this point.

\subsubsection{Refinement}

The \textsc{NewHorizon} simulation is run with the adaptive mesh refinement (AMR) code \textsc{ramses} \citep{Teyssier2002}. Within the high-resolution region, an initially uniform $4096^3$ cell grid is refined, according to a quasi Lagrangian criterion (when 8 times the initial total matter resolution is reached in a cell), with the refinement continuing until a minimum cell size of $\Delta x \sim34 \, \rm pc$ in proper units is achieved. Additional refinement is allowed at each doubling of the scale factor (at $a_{\rm exp}=0.1,0.2,0.4$ and $0.8$) in order to keep the resolution constant in physical units. The minimum cell size thus varies slightly between $\Delta x =27$ and 54~pc.  An additional super-Lagrangian refinement criterion is used to enforce the refinement of the mesh if a cell has a size shorter than one Jean's length and the gas number density is larger than $5\ \rm H\, cm^{-3}$ to better resolve the collapse of star-forming regions.

\subsubsection{Baryons}

Gas cooling is assumed to take place via H, He and metals, with an equilibrium chemistry model for primordial species (H and He) assuming collisional ionization equilibrium in the presence of a homogeneous UV background. Collisional ionization, excitation, recombination, Bremsstrahlung, and Compton cooling allow primordial gas to cool down to $\sim 10^4\ \rm K$, with metal-enriched gas able to cool further down to $0.1\, \rm K$ according to \citet{Sutherland1993} for temperatures above $\sim 10^4\rm\ K$ and according to \citet{Dalgarno72} below this value. The uniform UV background is switched on at $z = 10$, following \citet{Haardt1996}. UV photo-heating rates are reduced by a factor $\exp\left(-n/n_{\rm shield}\right)$, where $n_{\rm shield}= 0.01 \ \rm H\, cm^{-3}$. in order to account for self-shielding in optically thick regions \citep{Rosdahl2012}.

Star formation proceeds in regions with a hydrogen gas number density above $10\ \rm H\, cm^{-3}$ with stars particles forming with an initial mass of is $n_0 m_{\rm p} \Delta x^3=1.3\times 10^4 \ \rm M_\odot$ and following a Schmidt law ($\dot \rho_\star= \epsilon_\star {\rho_{\rm g} / t_{\rm ff}}$). $\epsilon_\star$ is a varying star formation efficiency \citep{Kimm2017} related to the star forming cloud properties through the cloud turbulent Mach number $\mathcal{M}=u_{\rm rms}/c_{\rm s}$ and virial parameter $\alpha_{\rm vir}=2E_{\rm kin}/E_{\rm grav}$.

Stellar feedback proceeds via Type II supernova assuming that each explosion initially releases kinetic energy of $10^{51}\ {\rm erg}$. SNe are assumed to explode instantaneously when a star particle becomes older than 5~Myr. The mass loss fraction from the explosions is 31 per cent, with the metal yield (mass ratio of the newly formed metals over the total ejecta) of 0.05. Both energy and momentum of the supernova is modelled, ensuring that the final radial momentum is accurately captured during the radiative phase of the supernova \citep{Kimm2014}. An addition to the final radial momentum from supernovae is also added in order to account for pre-heating of the ambient gas by OB stars \citet{Geen2015}.

\subsection{Galaxy magnitudes, colours and shapes}

\subsubsection{Galaxy magnitudes and colours}

\label{sec:mock_images}

For the purpose of measuring galaxy structural parameters other than shape measurements (Section \ref{sec:shapes}) as well as galaxy colours and magnitudes, we produce mock images in the rest frame using the \textsc{sunset} code (see \citealt{Kaviraj2017}), which implements dust attenuation via a dust screen model in front of each star particle. Spectral energy distributions (SEDs) are calculated from a grid of \citet[][BC03 hereafter]{Bruzual2003} simple stellar population (SSP) models interpolated to the age and metallicity of each star particle and assuming a Salpeter IMF. Rest-frame $g$ and $r$ band magnitudes are calculated by summing the resultant flux of each dust attenuated BC03 SED once it is convolved with the LSST $g$ and $r$  bandpass transmission functions \citep{Olivier2008}.

\subsubsection{Galaxy shapes}
\label{sec:shapes}

As we are required to calculate galaxy shapes at each snapshot, we neglect any treatment of dust attenuation in the interest of time, but otherwise the flux of each star particle is calculated in the same way. The major and minor axis at each snapshot is calculated in $xy$, $xz$ and $yz$ projections from the spatial distribution of star particles and their flux is obtained by first constructing the covariance matrix of their intensity-weighted central second-moment:

\begin{equation}
    \mathrm{cov}[I(x,y)]=\begin{bmatrix}
        Ix^{2}&Ixy\\
        Ixy&Iy^{2}\\
    \end{bmatrix},
\end{equation}

\noindent where $I$ is the flux of each star particle and $x$ and $y$ are their displacement from the barycentre. The major ($\alpha=\sqrt{\lambda_{1}/\Sigma I}$) and minor ($\beta=\sqrt{\lambda_{2}/\Sigma I}$) axes are obtained from the covariance matrix, where $\lambda_{1}$ and $\lambda_{2}$ are its eigenvalues and $\Sigma I$ is the total flux. 

\subsection{Identifying galaxies and producing merger trees}

\subsubsection{Producing merger trees}
\label{sec:merger_trees}

To identify galaxies we use the \textsc{HOP} structure finder \citep{Eisenstein1998}, applied to the distribution of star particles. Structures are identified if the local density exceeds 178 times the average matter density, with the local density being calculated using the 20 nearest particles. A minimum number of 50 particles is required to identify a structure. This imposes a minimum galaxy stellar mass of around $4\times 10^{5}$~M$_{\odot}$. We identify a total of 1017 galaxies in the simulation volume at $z=\basez$ of which 890 are dwarfs.

We produce merger trees for each galaxy using a base snapshot at a redshift of $z=\basez$. The time resolution of the merger trees is $\sim$15~Myr, enabling us to track in detail the main progenitors, assembly histories and mergers that each galaxy undergoes. Throughout the rest of this paper we refer to the higher mass galaxy of a merging pair as the \textit{primary} and its lower mass companion as the \textit{secondary}. We find good convergence between the merger histories of galaxies in the \textsc{NewHorizon} simulation and the lower resolution matching region in the Horizon-AGN simulation (see Appendix \ref{convergence}).

\subsubsection{Merger catalogues}
\label{sec:merger_catalogues}

Using these merger trees we produce a catalogue of mergers for each galaxy at the base snapshot. Following a similar procedure to \citet{Rodrigues2017}, we record the start of a merger, as well as measure stellar masses, gas content and merger mass ratio, at the snapshot just before the lower mass secondary companion begins to lose mass to the primary galaxy ($t_{\rm max}$), rather than at their coalescence in the merger tree. This is particularly important for accurately determining the merger mass ratio. Mergers are split into two classes: {\textit{major}} -- where the mass ratio ($R$) of the primary galaxy and its companion is greater than $R=1:4$ and \textit{minor} -- where $1:4>R>1:10$. Although the stellar mass accreted from so-called `mini mergers' with mass ratios smaller than 1:10 may be comparable to minor mergers, they are primarily important for the build up of stellar mass in stellar halo and have relatively little impact on the kinematics and properties of the central galaxy itself \citep{Arnaboldi2020,Schulze2020}.

The catalogues contain all the mergers that take place in the simulation. However, we only consider mergers where the primary galaxy of the merging pair is a main progenitor of one of the galaxies at the $z=\basez$ base snapshot. This is because we are interested specifically in their assembly (and including mergers that occur in galaxies before they have merged into the chain of one of the $z=\basez$ main progenitors would result in double counting of the same accreted mass). As such, when we present merger rates in Section \ref{sec:direct_accretion}, merger induced star formation budgets in Section \ref{sec:induced_star_formation} and other related quantities, they are not exactly equivalent to their global quantities as they do not take into account galaxies that underwent mergers before subsequently merging into the chain of one of the $z=\basez$ main progenitors. Additionally, by virtue having survived without merging into a more massive galaxies, low redshift dwarfs are, by nature, somewhat unusual objects compared to the sample of all dwarfs that have existed over cosmic time. For example at $z\sim3.5$, there are 2969 dwarfs identified within the simulation volume compared with only 1153 remaining dwarfs at $z\sim0.5$, meaning the vast majority of dwarfs that formed over cosmic time end up merging with more massive galaxies. Surviving dwarfs are therefore likely to have different assembly histories because dwarfs in denser environments are more likely to have merged into another galaxy by the present day compared with dwarfs in more isolated environments.

\begin{figure}
	\centering
    \includegraphics[width=0.45\textwidth]{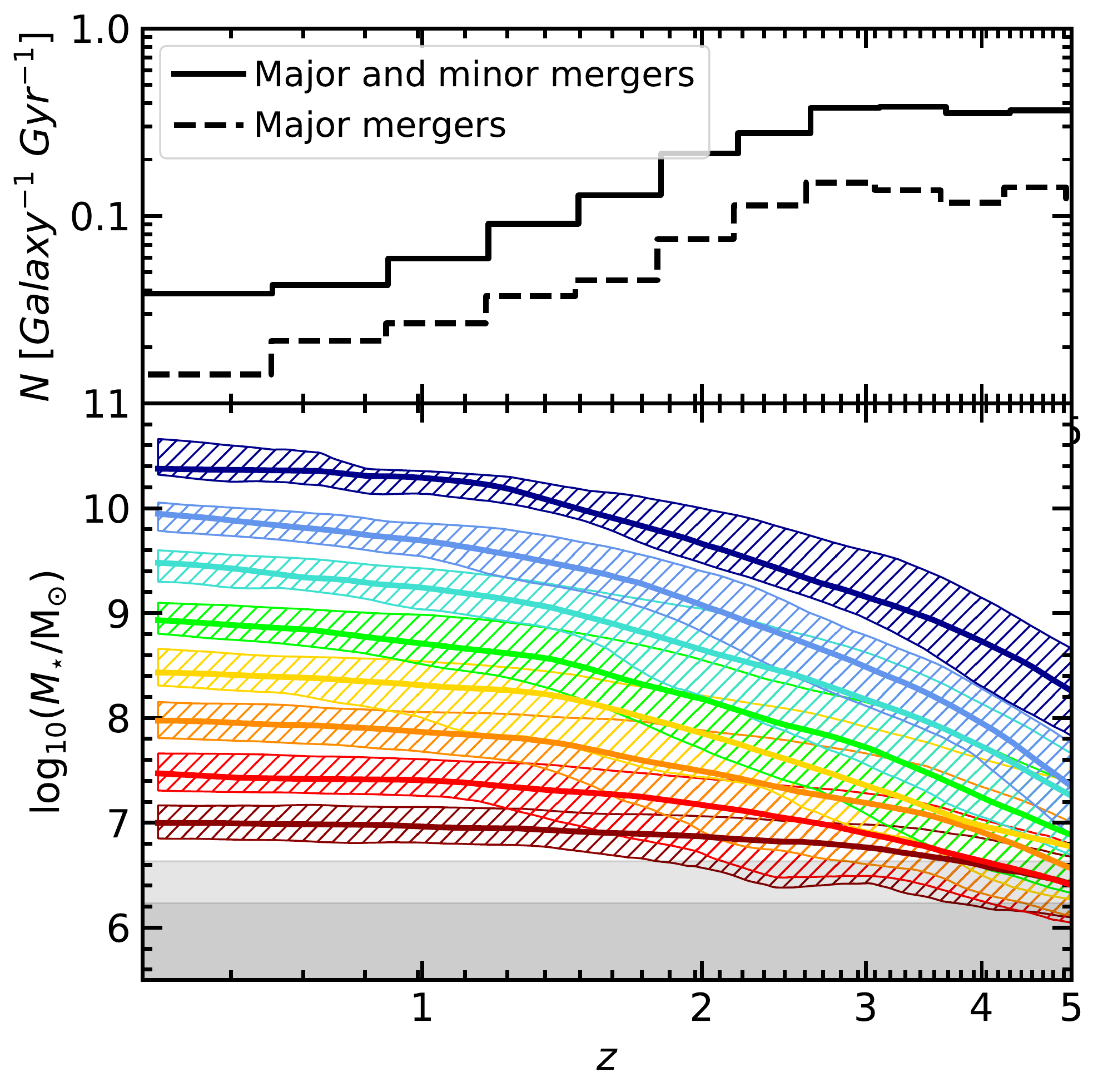}
    \caption{\textbf{Top}: Number of mergers per galaxy occurring in the \textsc{NewHorizon} simulation per Gyr as a function of redshift. \textbf{Bottom}: Median mass evolution of galaxies in bins of progenitor masses with central values of ${\rm log}_{10}(M_{\star}/{\rm M}_{\odot})$ 7, 7.5, 8, 8.5, 9, 9.5, 10 \& 10.5. The hatched region around each line indicates the $1\sigma$ scatter in their stellar mass. The dark shaded region indicates where minor mergers are undetected and the light shaded region indicates where only minor mergers are detected.}
    \label{fig:completeness}
\end{figure}

Fig \ref{fig:completeness} shows the merger rate (top panel) and median mass evolution of galaxies in bins of stellar mass of the $z=\basez$ main progenitor with central values of ${\rm log}_{10}(M_{\star}/{\rm M}_{\odot})$ 7, 7.5, 8, 8.5, 9, 9.5, 10 and 10.5. The light and dark grey regions show galaxy masses below which major and minor mergers are not detectable because the stellar mass of their secondary would be below the mass limit of $4\times 10^{5}$~M$_{\odot}$ imposed by the structure finder. A majority of major and minor mergers undergone by galaxies with masses greater than $10^{7}$~M$_{\odot}$ by $z=\basez$ are detectable at $z=4$ where the merger rate peaks. Major mergers are generally detectable in all mass bins up to $z=6$. In reality, the structure finder may not always correctly associate all of the tidal debris from a merging galaxy when the secondary galaxy has fewer than several hundred particles. We therefore limit our analysis to galaxies with progenitor stellar masses larger than $10^{7.5}$~M$_{\odot}$ in order to ensure that $R>1:10$ mergers can be both reliably identified and correctly dealt with by the structure finder at high redshift (See Appendix \ref{convergence} for further discussion).

We do not consider galaxies with progenitor masses lower than $10^{7.5}$~M$_{\odot}$ in our larger analysis, although we do show lower mass bins for illustrative purposes, since they are largely reliable at lower redshifts. In any case the inclusion of exclusion of galaxies of these masses does very little to alter our conclusions as they account for a very small proportion of the stellar mass budget and undergo very few mergers.

\subsection{Morphological disturbances}

\label{sec:disturbances_def}

\begin{figure}
	\centering
    \includegraphics[width=0.45\textwidth]{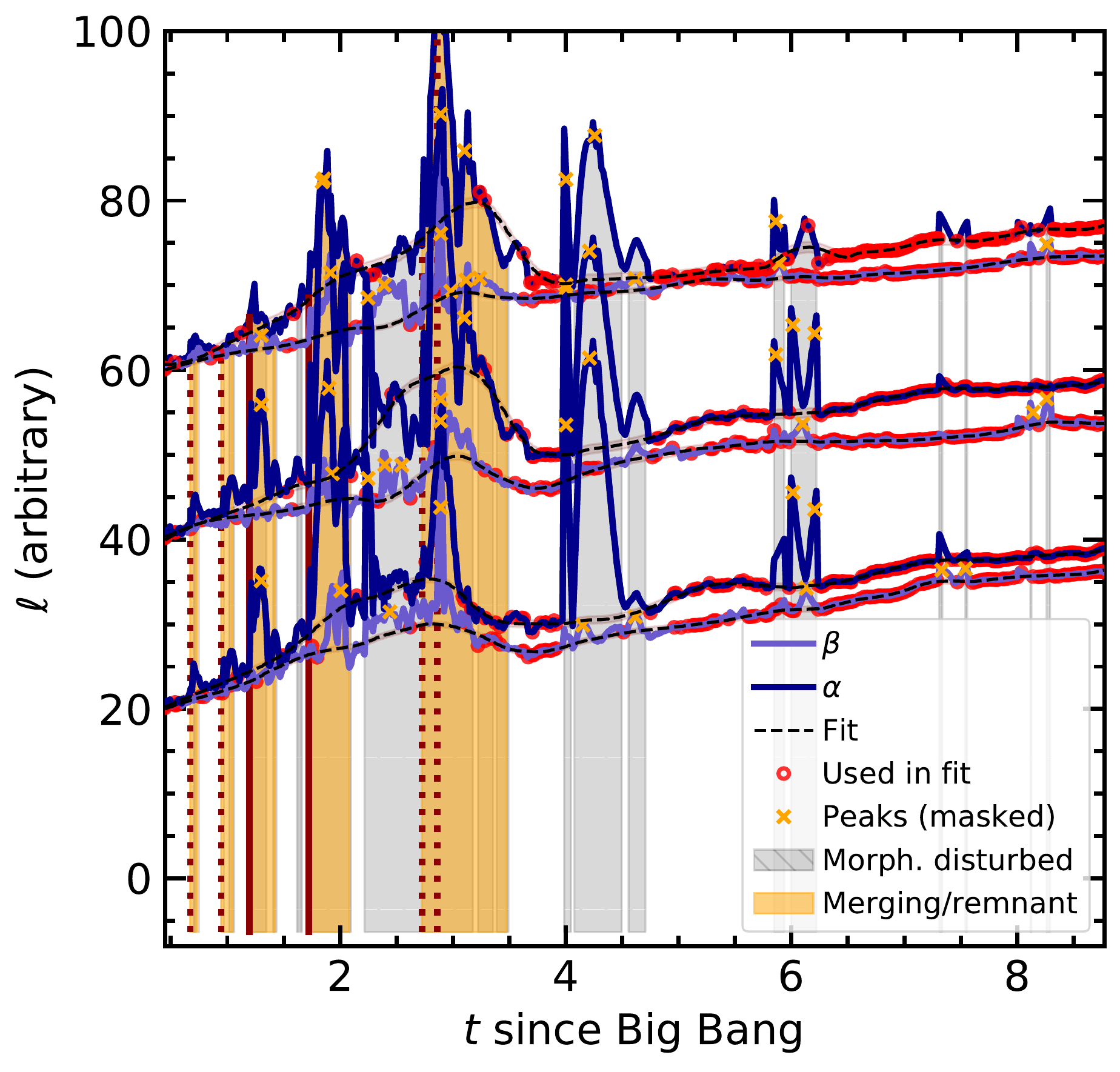}
    \caption{Plot illustrating our method for determining when an object is morphologically disturbed. Light and dark blue lines show the major and minor axes of the galaxy in 3 different projections (with arbitrary normalisation), red points show the points used to fit the continuum and the black line indicates the fit with the shaded region indicating the standard deviation of the residual. Yellow crosses show the detected peaks which were masked out from the fit and grey and orange shaded regions show the times that the galaxy is determined to be disturbed and disturbed as the result of a merger respectively.}
    \label{fig:fit_example}
\end{figure}

In order to determine whether a galaxy is morphologically relaxed or disturbed, we consider the variation in the length of the major and minor axes, $\alpha$ and $\beta$, based on the $r$-band surface brightness profile of each galaxy over time. We specify criteria for whether a galaxy is at rest or disturbed based on the assumption that the size and shape of a galaxy that is evolving secularly and undergoing no external perturbations will evolve gradually and smoothly over time, whereas a galaxy that is subject to some external disturbance will depart from this gradual evolution before returning after a period of time. 

We first smooth the axis length evolution vs time, and mask out any peaks:

\begin{enumerate}
\item \noindent Perform a boxcar average with a width of $\sim 75$~Myr (5 snapshots) in order to smooth the signal for the evolution of the major and minor axis length in each projection.\\ 
\item \noindent Using the smoothed signal, locate all local maxima with a prominence larger than the \textit{local} standard deviation (computed as the boxcar average with a width of 5 snapshots) and a width larger than 3 snapshots ($\sim 45$~Myr) using the scipy function {\tt scipy.signal.find\_peaks } \citep{2020SciPy} (yellow crosses in Fig \ref{fig:fit_example}).\\
\item \noindent Using the unsmoothed data for both axes in each projection, select a sub-sample of points by first discarding all points that coincide with detected peaks and then selecting 25 per cent of the remaining points with the smallest local gradient (red open circles in the top panel of Fig \ref{fig:fit_example}).\\
\item \noindent Use a Savitzky-Golay filter \citep{Savitzky1964} (implemented using the scipy function {\tt scipy.signal.savgol\_filter} \citealt{2020SciPy}) on each sub-sample with a window length of 1~Gyr (chosen to be at least one orbital timescale) and a polynomial order of 5 to obtain a smooth `\textit{continuum}' fit to the data (dotted black lines in Fig \ref{fig:fit_example}).
\end{enumerate}

We then calculate $\sigma_{\rm local}$, which we define as the local standard deviation of the residual between the smoothed continuum fit and the sub-set of points used to obtain the fit as a boxcar average with a width of 5 snapshots. A galaxy is considered to be \textit{morphologically disturbed} whenever the actual value of the axis length deviates by more than 3.5 times $\sigma_{\rm local}$. This threshold is chosen based on visual inspection of high resolution mock images and to approximately match the major merger timescales obtained from non-parametric morphological indicators (Fig \ref{fig:timescales}). In Appendix \ref{sec:test}, we explore how changing this threshold or using a fixed threshold affects merger durations.

\begin{figure}
	\centering

    \includegraphics[width=0.45\textwidth,trim=0 1.2cm 0 0]{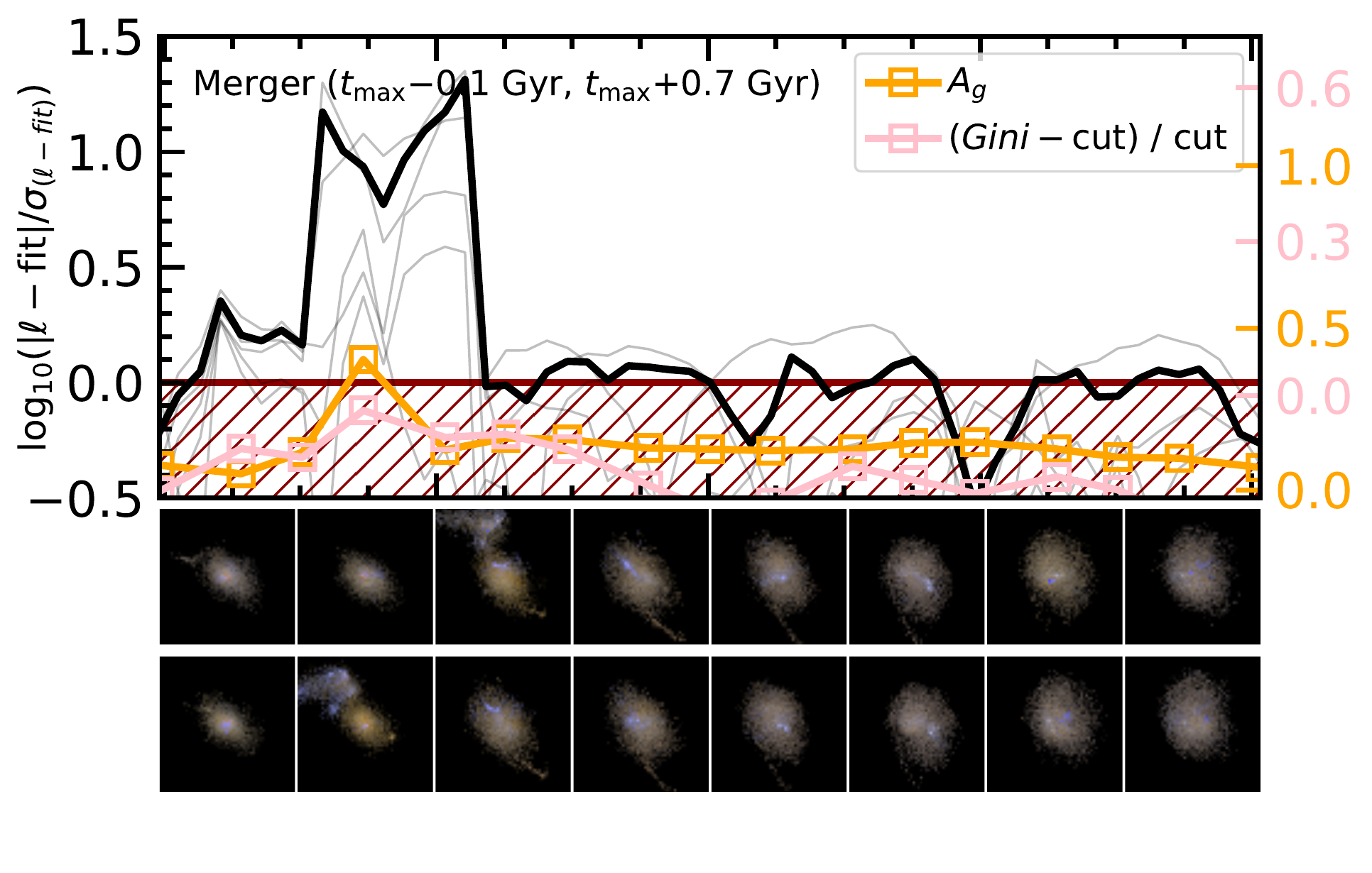}
    \includegraphics[width=0.45\textwidth,trim=0 1.2cm 0 0]{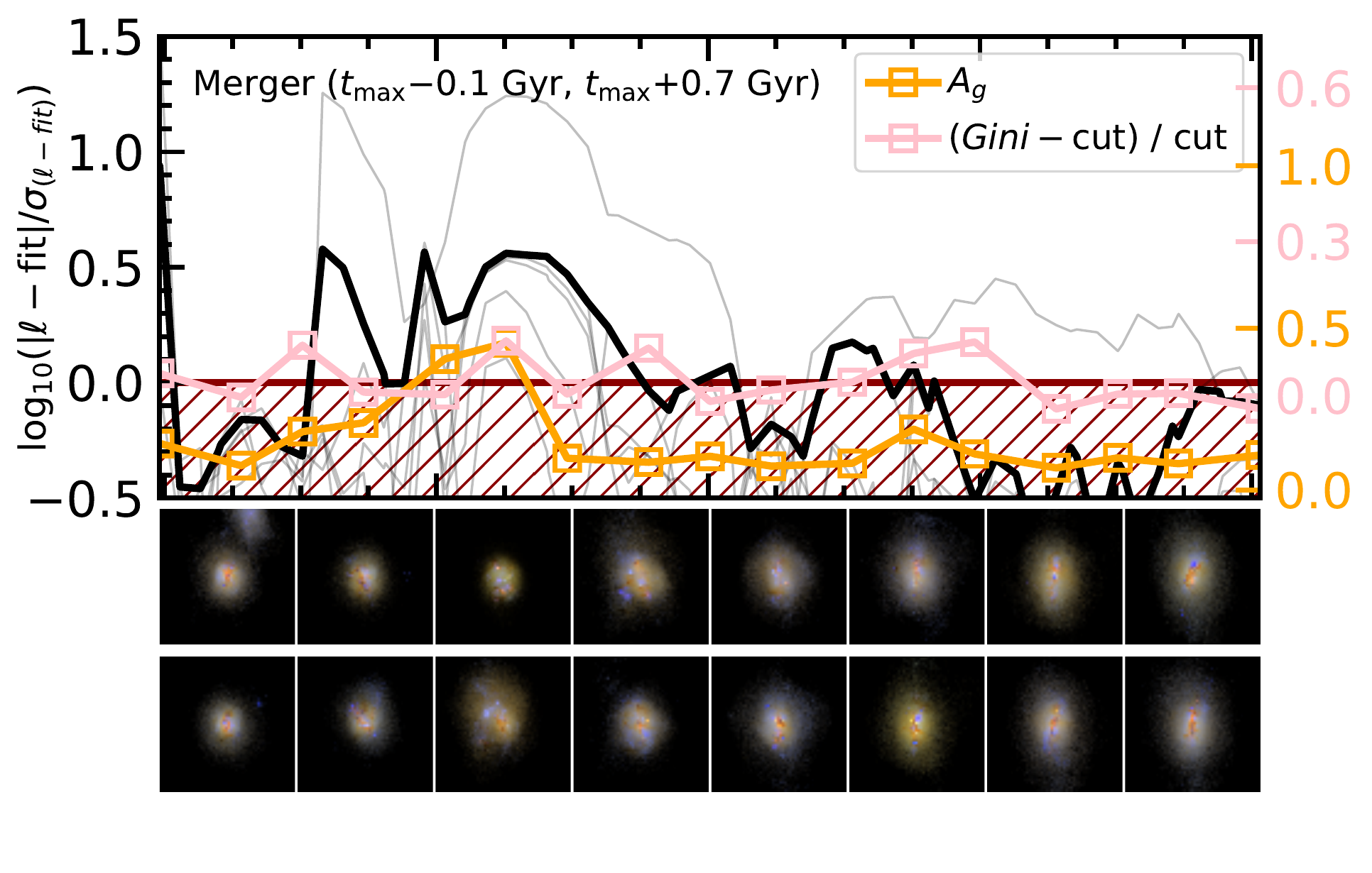}
    \includegraphics[width=0.45\textwidth,trim=0 1.2cm 0 0]{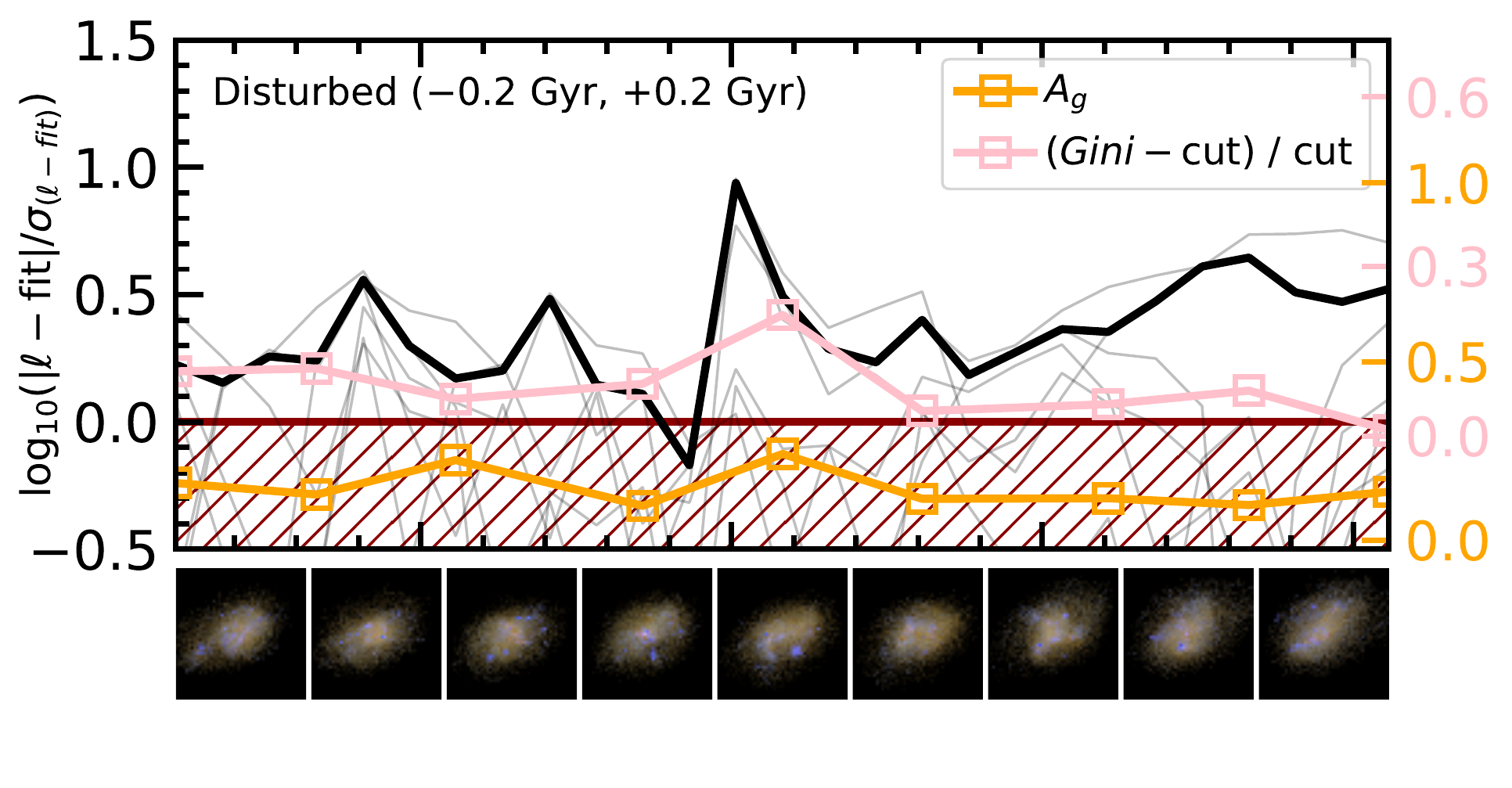}
    \includegraphics[width=0.45\textwidth,trim=0 1.2cm 0 0]{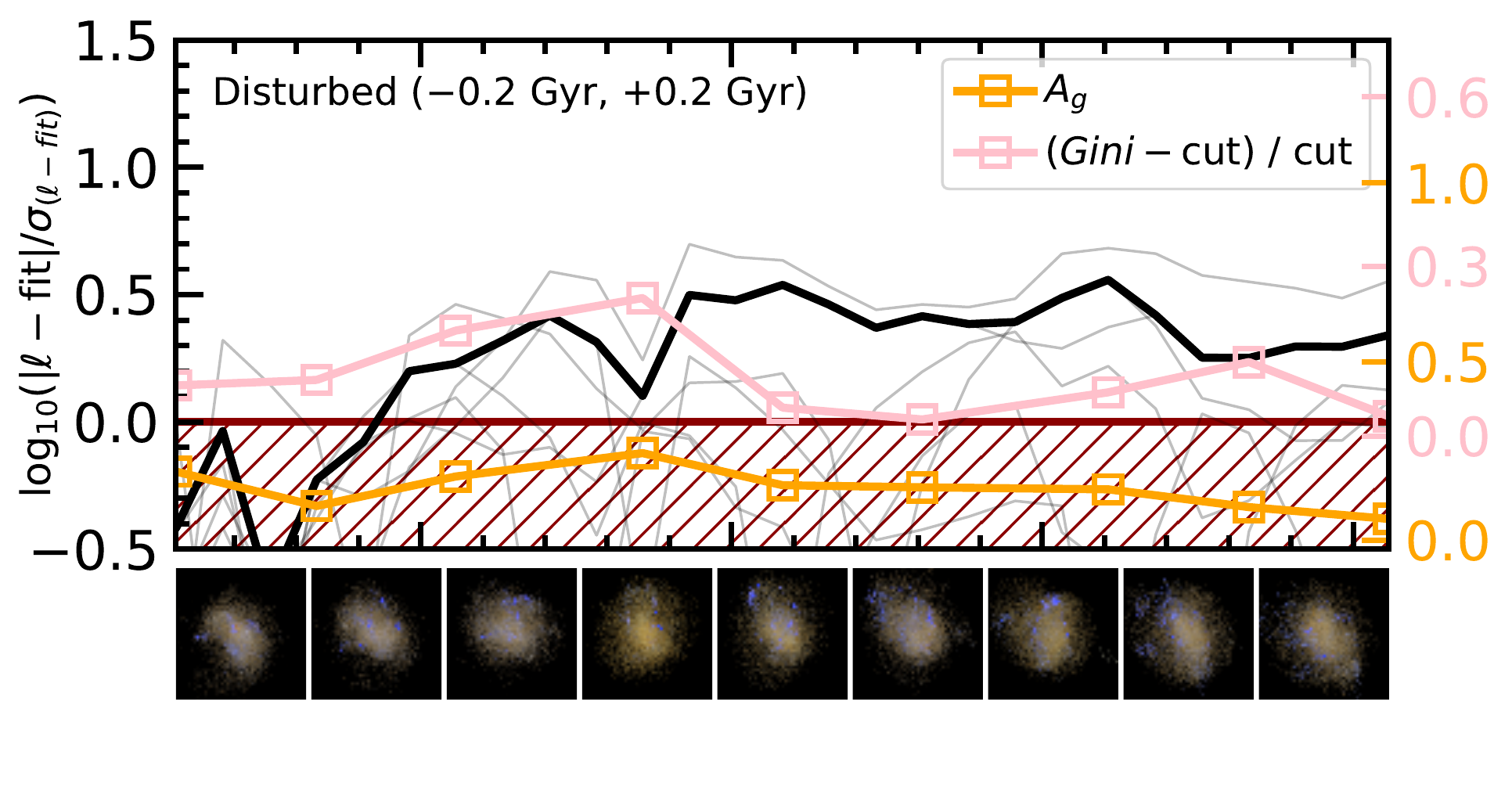}

    \caption{The top part of each panel shows the residual of the axis lengths and continuum for each axis and second highest value of the residual (thin and thick black lines respectively), $A_{g}$ (yellow) and the distance of the $Gini$ coefficient above the $Gini$-$M_{20}$ cut (pink). $y$-axes are scaled so the red hatches enclose a region where the galaxy is relaxed according to our criteria, $A_{g}$ and $Gini$-$M_{20}$. Bottom panels show $u$-$r$-$z$ images in chronological order (top to bottom / left to right), the position of each image corresponds roughly to the time in the plot above. The top two panels show mergers between $100$~Myr and $700$~Myr before and after $t_{\rm max}$. The bottom two panels show morphologically disturbed galaxies for $\pm200$~Myr.}
    \label{fig:mock_example}
\end{figure}

Fig \ref{fig:mock_example} shows two example merger sequences and two example disturbed galaxies with $u$-$r$-$z$ false colour images as well as the evolution of the residual of the axis lengths and the continuum, the shape asymmetry, $A_{g}$ and distance of the $Gini$ coefficient above the $Gini$-$M_{20}$ cut (see Appendix \ref{sec:merger_test}). Note that for both mergers, tidal features are visible following the merger but the solid black line falls below the threshold required for the galaxy to be considered morphologically disturbed (also true of asymmetry and $Gini$-$M_{20}$). This is because our method uses the axis lengths of the galaxy and is therefore mostly sensitive to the bright central part of the galaxy rather than extended low surface-brightness features.


In order to classify, mergers and disturbed galaxies, we consider two possible types of morphological disturbance -- morphological disturbances that are not accompanied by a merger \citep[e.g. those due to fly-bys][]{Miller1986} and merger remnants:

\begin{itemize}
    \item \textit{Morphological disturbance / interaction:} wherever the lengths of at least two axes in any combination of projections deviate from the continuum by more than 3.5 times $\sigma_{local}$. Grey regions in Fig \ref{fig:fit_example} indicate where the galaxy is considered morphologically disturbed. This may be the result of any processes that produce variance in the shape of a galaxy. For example disruption to the galaxy as a result of a close encounter with another galaxy that does not end in a merger, tidal forces or other environmental mechanisms. See Appendix \ref{sec:PI_test} for discussion of the main processes driving this morphological disturbance.\\
    
    \item \textit{Merger / merger remnant:} wherever at least two axes in any combination of projections deviate from the continuum fit to the axis length evolution by more than 3.5 times $\sigma_{\rm local}$ coincident with a merger of a mass ratio of at least $1:10$. The start of the merger must be within 3 snapshots (45~Myr) of a snapshot where the galaxy is disturbed and the merger remnant is considered relaxed once the deviation from the continuum fit in at least 5 of the axes from the three projections remains below 3.5 times $\sigma_{\rm local}$ for more than 3 snapshots. Yellow regions in Fig \ref{fig:fit_example} indicate where the galaxy is considered to be merging or a merger remnant and red lines indicate the times where a merger has taken place (minor mergers as dotted red lines and major mergers as solid red lines).\\
\end{itemize}

For each merger, we also calculate $\tau_{\rm merger}$, the time between the beginning of the merger, as defined by $t_{\rm max}$ and the end of the merger, defined as the point at which the deviation in fewer than two axes are above 3.5 times  $\sigma_{\rm local}$ for more than 3 snapshots (ending at the first point that they were all below 3.5 times  $\sigma_{\rm local}$, not following 3 additional snapshots). $t_{\rm max}$ generally occurs a few snapshots (a few tens of Myrs) before the coalescence in the merger tree, meaning that $\tau_{\rm merger}$ mostly describes the time that the merger remnant remains disturbed.

We detect axis variations coincident with almost all $R>1:10$ mergers (96 per cent). The remaining 4 per cent of $R>1:10$ mergers for which we do not detect any axis variations above 3.5~$\sigma_{\rm local}$ are discarded and not considered in our analysis.

In principle deviations in the axis lengths of disturbed galaxies may also be the result of clumpy star formation. This is likely to be more common in lower mass galaxies, where star formation tends to take place in localised regions \citep[e.g.][]{Sanchez2015}. We consider this possibility by employing a perturbation index \citep[PI;][]{Byrd1990} which quantifies the environmental tidal field due to objects in the vicinity of the galaxy in question. PI is defined according to \citet{Choi2018} at low redshift where the star formation enhancement in morphologically disturbed galaxies is highest relative to undisturbed galaxies (Section \ref{sec:merger_enhancement}):

\begin{equation}
\label{eqn:PI}
{\rm PI} = \sum_{i}\left ( \frac{M_{i}}{M_{tot}}\right ) \left ( \frac{R_{\mathrm{eff}}}{D_{i}} \right )^{3},
\end{equation}

\noindent where $M_{tot}$ is the total mass of the galaxy in question, $M_{i}$ is the mass of the $i$th perturbing galaxy, $R_{\mathrm{eff}}$ is the effective radius and $D_{i}$ is the distance from the $i$th perturbing galaxy.

In order to test whether star formation can produce morphological disturbances large enough to be detected by our method, we consider the average enhancement of the specific star formation rate (sSFR) in disturbed (non-merging) galaxies, compared with their undisturbed counterparts at fixed PI. Any enhancements above the undisturbed control sample would be an indication that star formation plays some role in producing detectable morphological disturbances, since higher star formation rates should produce larger disturbances independent of the external tidal field. We do this by constructing a sub-sample of disturbed galaxies with the same PI distribution as the undisturbed galaxies at the lowest redshift snapshots that we have analysed (between $z=0.48$ and $z=0.52$) and comparing the average sSFR of these two samples. We obtain a slight depression of $0.94\pm_{0.06}^{0.10}$ times for the dwarf regime ($7.5<M_{\star}/{\rm M_{\odot}}<9$), consistent with there being no enhancement. For the full sample of disturbed galaxies, we obtain an average sSFR enhancement of 3 times and the enhancement in PI is around 2 times. This indicates that internal star formation processes are responsible for a negligible fraction of the morphological disturbances that we detect using our method and are instead the result of genuine interactions.

\section{The frequency and durations of galaxy morphological disturbances}
\label{sec:freq_disturbances}

In this section, we examine how mergers and interactions drive morphological disturbances in galaxies over time. In Section \ref{sec:disturbance}, we investigate how the mechanisms that drive morphological disturbances evolve with stellar mass and in Section \ref{sec:timsecales}, we investigate the duration of these morphological disturbances.

\subsection{Morphological disturbances as a function of stellar mass}
\label{sec:disturbance}

In order to determine the fraction of galaxies that exhibit morphological disturbances as a function of stellar mass, we calculate the time that each galaxy is morphologically disturbed or merging as a result of major or minor mergers as a fraction of their lifetime (i.e. beginning once they have formed $4\times10^{5}$~M$_{\odot}$ of stellar mass and are therefore detected as structures to the base snapshot at $z=\basez$).

\begin{figure}
	\centering
    \includegraphics[width=0.45\textwidth]{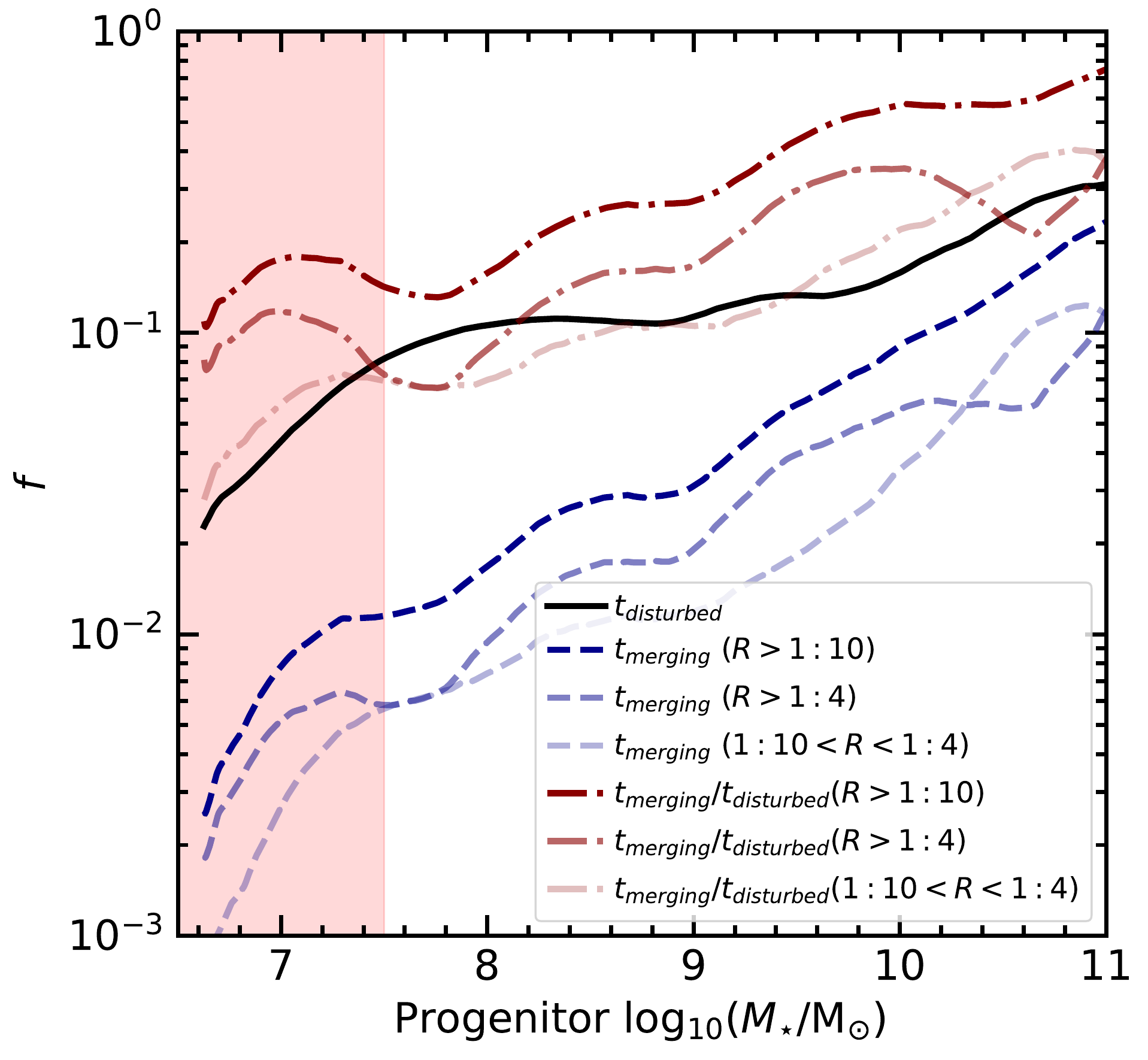}
    \caption{Average fraction of their lifetime that galaxies spend in a disturbed state ($t_{\rm disturbed}$, solid black line), disturbed as a result of mergers of different mass ratios ($t_{\rm merging}$, blue dashed lines) and the ratio of the two (red dot-dashed lines) as a function of their progenitor stellar mass at $z=\basez$. The dark red lines indicate these values for all mergers with $R>1:10$, the lighter lines indicate major mergers only and the lightest lines indicate minor mergers only. We do not consider any galaxies with progenitor masses in the shaded red region, as they can become too poorly resolved at high redshift ($z\sim5$).}
    \label{fig:morphological_disturbance}
\end{figure}

The solid black line in Fig \ref{fig:morphological_disturbance} shows the average fraction of their lifetime that galaxies are in a morphologically disturbed state. Galaxies of all stellar masses spend significant amounts of time (between 10 and 30 per cent) in a morphologically disturbed state on average, with the most massive galaxies in the sample having somewhat enhanced fractions. Blue lines indicate the integrated fraction of time that galaxies are morphologically disturbed as a result of major and minor mergers and red lines indicate the fraction of all morphological disturbances that are driven by mergers. As the blue lines show, the most massive galaxies spend a significant fraction of their lifetime as mergers or merger remnants, whereas for galaxies of $10^{7.5}~{\rm M}_{\odot}$, this value is around 1 per cent. Mergers are responsible for less than 20 per cent of morphological disturbance seen in dwarf galaxies ($M_{\star} < 10^{9}~{\rm M}_{\odot}$) whereas mergers quickly come to dominate at stellar masses greater than $10^{9.5}~{\rm M}_{\odot}$, accounting for close to 100 per cent in the highest mass bins. If we vary the threshold for morphological disturbance from 3.5~$\sigma_{\rm mult}$, we do not see any significant variation in this fraction.

The increase in the importance of mergers for producing morphological disturbances at higher masses is largely a consequence of the fact that the merger rate increases towards higher masses, with massive galaxies having around 10 times more mergers than dwarfs (see Section \ref{sec:direct_accretion}). The red region indicates progenitor masses that we exclude from our analysis for reasons outlined in Section \ref{sec:merger_catalogues}. We note that if we restrict our analysis only to lower redshifts where we know all galaxies are well enough resolved for all mergers to be detectable ($z < 2$) we do not see the same sudden drop in merger fraction for progenitor stellar masses smaller than $10^{7.5}~{\rm M}_{\odot}$, indicating that merger rate continues to evolve smoothly below $10^{7.5}$~${\rm M_{\odot}}$.

\begin{figure}
	\centering
    \includegraphics[width=0.45\textwidth]{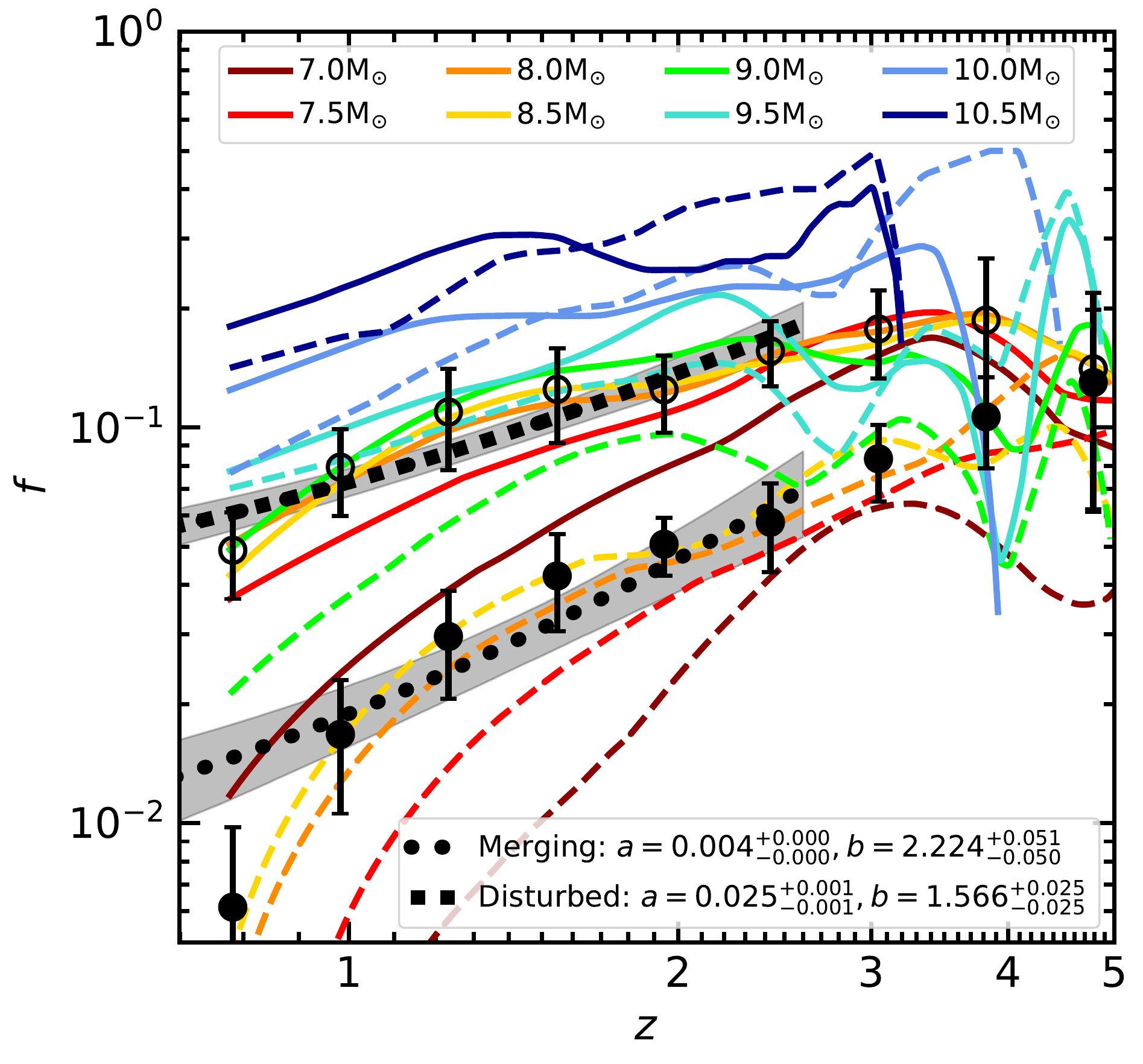}
    \caption{Fraction of galaxies that are morphologically disturbed at a given redshift (solid lines) and morphologically disturbed as the result of mergers with mass ratios greater than $1:10$. Colours indicate the $z=\basez$ progenitor mass and thick dotted lines indicate the average fractions across all masses. Error bars with open and filled circles indicate average fractions for disturbed and merging fractions for the dwarf regime ($7.5<{\rm log_{10}}(M_{\star}/{\rm M_{\odot}})<9$), lines dotted with squares and circles indicate the best fitting parameters to the function $f=a(1+z)^{b}$ using these data points and shaded regions indicate the $1\sigma$ uncertainty. Error bars and model uncertainties are both multiplied by 10 for legibility.}
    \label{fig:morphological_disturbance_z}
\end{figure}

Similar to Fig \ref{fig:morphological_disturbance}, Fig \ref{fig:morphological_disturbance_z} shows the fraction of galaxies in different mass bins that are morphologically disturbed \textit{at a given redshift}. Solid lines indicate all morphologically disturbed galaxies and dashed lines indicate only merger induced morphological disturbances. At any given redshift, more massive galaxies are most likely to be merger remnants or undergoing a merger. However, when looking at all disturbed galaxies, regardless of the source of the morphological disturbance, the same trend is not evident, with galaxies of all stellar masses exhibiting similar evolution. The evolution of the merger fraction in high mass galaxies is in agreement with a previous study by \citet[][]{Kaviraj2015a} using the Horizon-AGN simulation, which shows that the merger fraction of high mass galaxies ($M_{\star}>10^{10}$~M$_{\odot}$) does not evolve strongly with redshift between $z=4$ and $z=1$.

The fraction of mergers and merger remnants is relatively high at $z>3$; greater than 5 per cent for dwarf galaxies and around 30 per cent for the most massive galaxies. This corresponds well to major merger fractions obtained from observational studies of peculiar galaxies, which typically find values of between 25 and 50 per cent at high redshift \citep{Conselice2003,Lotz2006,Bluck2012,Cibinel2019}. Towards $z=1$, where morphological merger fractions derived from asymmetry and double nuclei in high resolution imaging \citep[e.g.][]{2009LopezSanjuan,Lackner2014} are around 10 per cent for massive galaxies, we again find good agreement for high masses (where the merger fraction is approximately equal to the morphologically disturbed fraction). 

In all mass bins, merger fractions decrease towards low redshift as the merger rate drops. The evolution in the disturbed fraction is not as strong and decreases more gently towards low redshifts. In order to quantify this evolution, we consider the slope of the redshift evolution of the merging and disturbed fractions for dwarf galaxies in the redshift range $z<2.5$. We fit the merger and disturbed fractions using the functional form $f=a(1+z)^{b}$, obtaining, for the merger fraction, an index of $b=2.224\pm0.050$ and for the disturbed fraction, an index of $b=1.566\pm0.025$.

As we have shown in this section, it may only be reasonable to assume that disturbed morphologies are a good proxy for past merger activity in the high mass regime, since this is where mergers dominate over other drivers of morphological disturbances. Like higher mass galaxies, dwarf galaxies are morphologically disturbed for a significant fraction of their lifetimes (10 per cent to 20 per cent) and mergers make an increasingly small contribution to these disturbances towards lower masses, meaning this assumption is not valid for stellar masses below  $10^{9}{\rm~M_{\odot}}$. Apparently merging dwarf galaxies are instead more likely to be disturbed as a result of processes unrelated to mergers.

It is interesting to consider these results in the context of future studies of galaxies in the dwarf regime. While we can model these objects and explore some of their properties in the local Universe, dwarf galaxies are not yet accessible to contemporary instruments at intermediate or high redshifts. Enabled by instruments like JWST and LSST, studies relying on using galaxy morphology to infer merger activity or merger rates in dwarf galaxies will require considerable care in disentangling or correcting for the large fraction of galaxies that appear morphologically disturbed, but are not merging. This will be particularly important when attempting to identify high redshift mergers and post mergers morphologically \citep[e.g.][]{2019Mantha} as well as in interpreting previous work at lower redshifts \citep[e.g.][]{Paudel2018,Kaviraj2019} given the large fraction of disturbed objects that could be confused for mergers. In Appendix \ref{sec:merger_test}, we show how our morphological disturbance measure correlates with non-parametric shape asymmetry and concentration ($A_{g}$-$C$) and $Gini$-$M_{20}$ coefficients used to identify mergers. We see that, while major mergers, minor mergers and disturbed galaxies can be disentangled from the undisturbed population, they occupy similar regions of parameter space for both $A_{g}$-$C$ and $Gini$-$M_{20}$ at JWST-like resolutions. However, none of these measures, including our measure of morphological disturbance (Section \ref{sec:disturbances_def}) are very sensitive to merger relics like tidal tails since they carry only a few per cent of the total stellar mass of the galaxy \citep{Kim2012}. It is therefore likely that given well resolved objects with deep enough imaging, merging galaxies could still be identified by visual inspection of their low surface brightness features.

\begin{figure}
	\centering
    \includegraphics[width=0.45\textwidth]{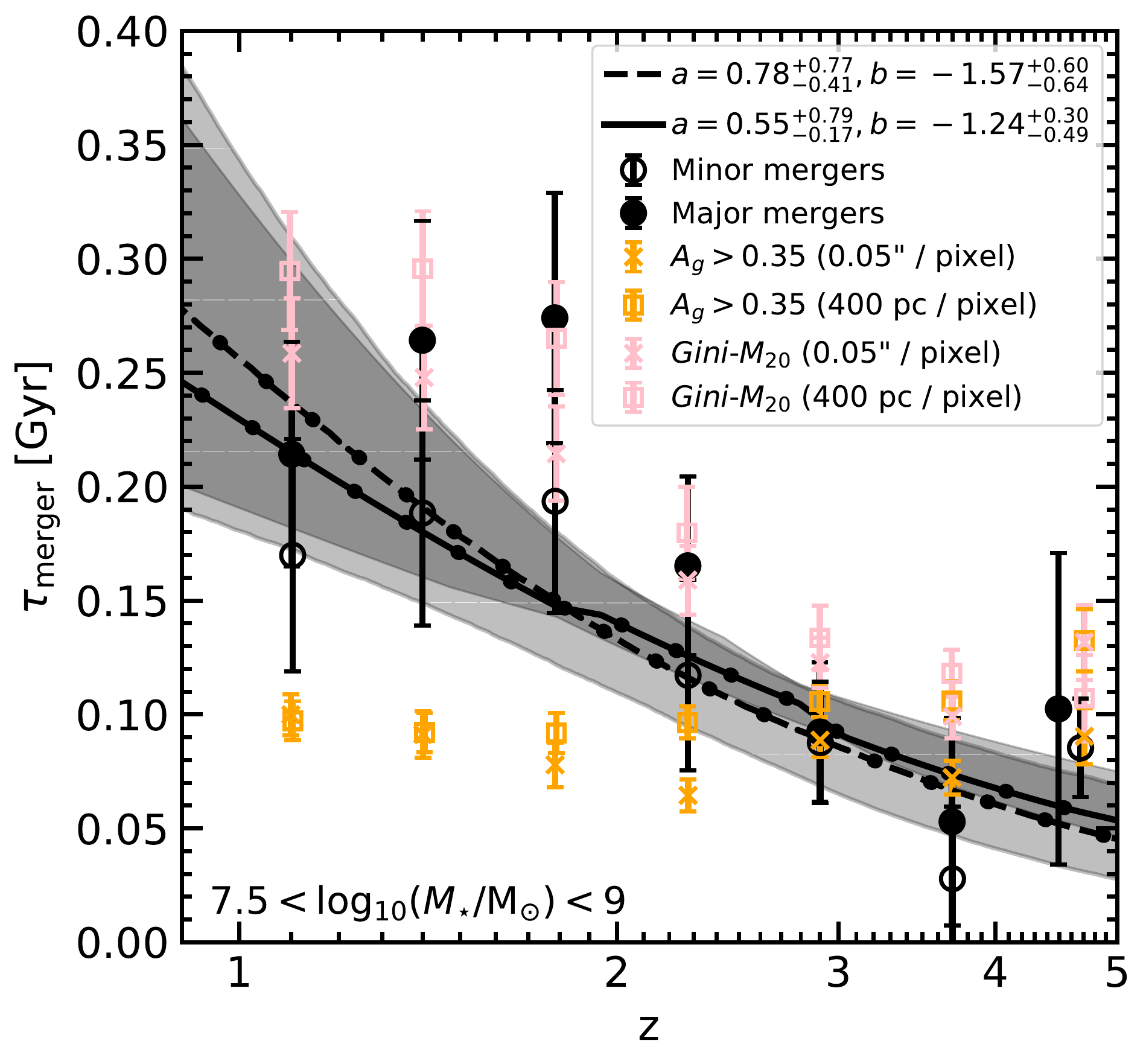}
    \caption{Median observable duration of major (filled points) and minor mergers (open points) as a function of redshift for the dwarf regime ($7.5<{\rm log_{10}}(M_{\star}/{\rm M_{\odot}})<9$). Solid and dashed lines show a fit to the data ($\tau_{\rm merger} = a(1+z)^{b}$) and filled regions indicate the $1\sigma$ uncertainties. Orange symbols with error bars indicate asymmetry timescales for major mergers only derived from mock images at a fixed angular scale for 0.05" / pixel (crosses) and 400~pc / pixel (open squares).}
    \label{fig:timescales}
\end{figure}

\begin{figure*}
	\centering
    \includegraphics[width=0.95\textwidth]{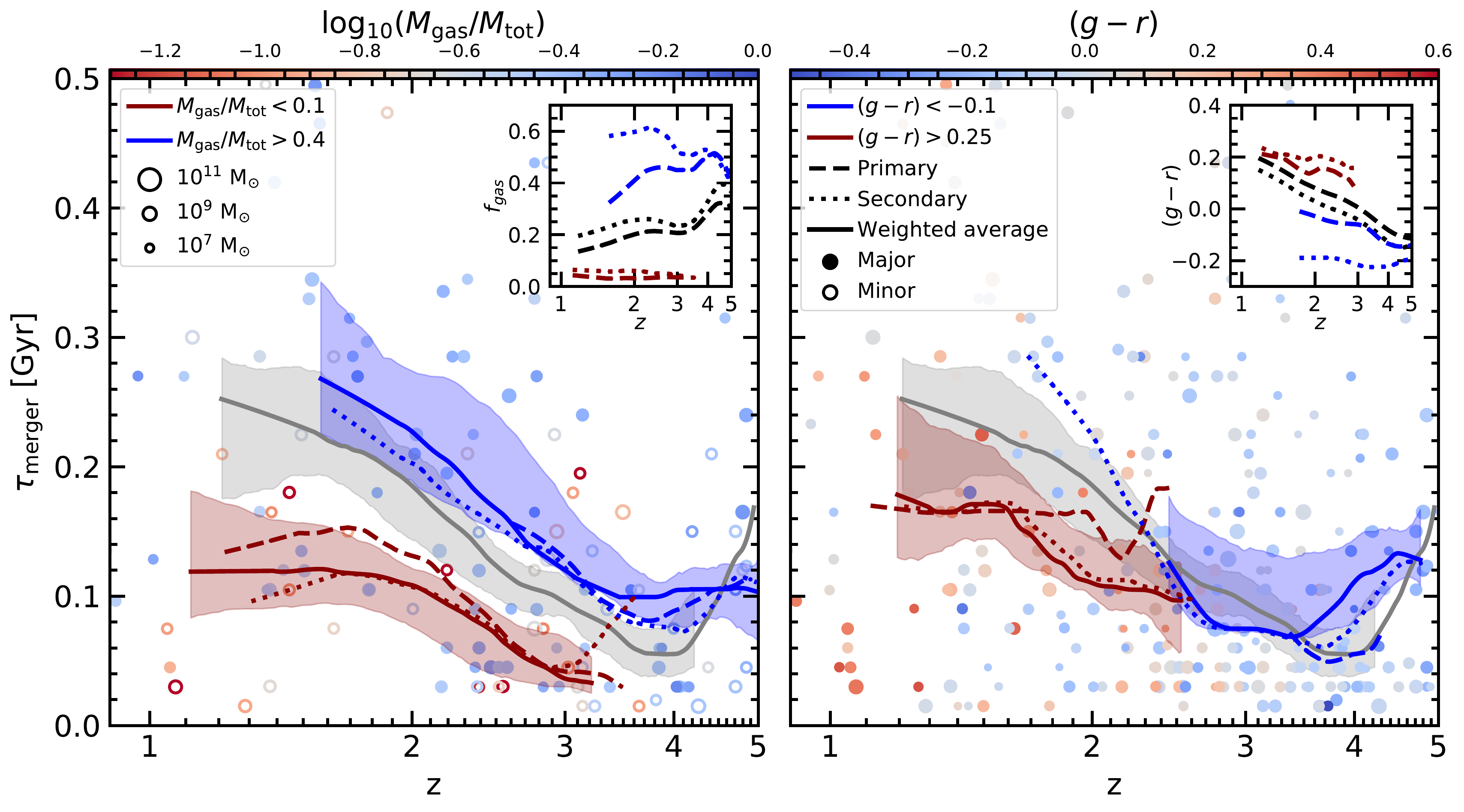}
    \caption{Scatter plots showing merger duration as a function of redshift. The size of each point corresponds to the stellar mass of the host galaxy (as indicated in the legend), with open circles corresponding to minor mergers and filled points corresponding to major mergers.  \textbf{Left}: Points are colour coded by the combined gas fraction of the merging pair. Solid, dashed and dotted lines show the median merger timescale when we split mergers into sub-samples based on the combined gas fraction of the pair, the gas fraction of the primary and the gas fraction of the secondary respectively. The grey line indicates the median for the full sample. Shaded regions indicate $1\sigma$ uncertainties. For clarity, only the uncertainties for the combined gas fraction are shown but uncertainties are similar for the other samples. The inset plot indicate the evolution of the average gas fraction of the primary and secondary as a function of redshift. \textbf{Right}: Points are colour coded by the luminosity weighted average colour of the merging pair. Solid, dashed and dotted lines show the trend for the luminosity weighted average colour of the pair, the colour of the primary and the colour of the secondary respectively. The inset plot indicate the evolution of the colour of the primary and secondary as a function of redshift.}
    \label{fig:timescales:gas}
\end{figure*}

\subsection{Merger duration and observability}
\label{sec:timsecales}

In this section we consider the duration over which merger and merger remnants remain morphologically disturbed. Galaxies are considered to be merging based on the definition in Section \ref{sec:disturbances_def} -- each merger begins at $t_{\rm max}$ (i.e. once the pair have begun to exchange mass) and is considered to have ended (i.e. the pair have coalesced and the remnant has relaxed) once the merger remnant does not deviate significantly from the continuum for more than 3 snapshots. We consider only mergers where the primary galaxy is not morphologically disturbed before the merger begins and which do not undergo another merger until after the merger remnant has relaxed (54 per cent of all mergers).

Fig \ref{fig:timescales} shows average merger durations, $\tau_{\rm merger}$, in the dwarf regime ($M_{\star}<10^{9}$~M$_{\odot}$) as a function of redshift. Open circles show merger durations for minor mergers and filled circles show merger durations for major mergers based on the median value within bins of width 0.5 in redshift and with error bars indicating the $1\sigma$ error. The duration of mergers increases towards lower redshifts. Although major and minor mergers remnants have similar durations at high redshifts ($z<3$), major mergers remnants take a similar amount of time to relax than minor mergers at low redshifts (although merger relics like tidal tails, shells and distortions are likely to persist longer in systems that have undergone a minor merger e.g. \citealt{Eliche2018}).

We perform a power law fit ($\tau_{\rm merger} = a(1+z)^{b}$) to the data for dwarf minor and major mergers, which are shown as dotted and solid grey lines respectively. Filled regions indicate the $1\sigma$ uncertainty from 10,000 bootstraps. The best fitting values and their errors are $a = 0.78^{+0.37}_{-0.41}$, $b = -1.57^{+0.79}_{-0.17}$ for minor mergers and $a = 0.55^{+0.39}_{-0.37}$,  $b = -1.24^{+0.30}_{-0.49}$ for major mergers.

For comparison, we also present average observability timescales for major merger remnants using a shape asymmetry threshold of $A_{g}=0.35$ and cuts in $Gini-M_{20}$. Our method is described in detail in Appendix \ref{sec:merger_test}. Since we find that both the $A_{g}$ and  $Gini-M_{20}$ measures can drop temporarily below the merger threshold, for example as a result of the projection of the merging secondary in front of or behind the primary, the timescales are simply calculated to be the total time that the merger remnant is above the threshold within a 1~Gyr window following $t_{\rm max}$, not the time before the merger remnant drops below the threshold for the first time. Out of a total of 626 major mergers between $z=1$ and $z=6$, 593 and 534 are detectable at any point (95 and 85 per cent) using the $A_{g}$ and $Gini-M_{20}$ methods respectively. Of these, 65 and 88 per cent are below the $A_{g}=0.35$ and $Gini-M_{20}$ threshold respectively 100~Myr before $t_{\rm max}$, meaning they would not have been classified as a merger before the merger began. Mergers that do not meet this criteria are discarded since their asymmetry may be the result of a previous interaction.

We present the asymmetry and $Gini-M_{20}$ observability timescales using mock images with fixed angular size and fixed physical size. indicated in Fig \ref{fig:timescales} as orange and pink open squares and crosses respectively. At fixed physical scale, asymmetry timescales remain relatively flat over time ($100-200$~Myr), while the $Gini-M_{20}$ timescale increases towards lower redshift. At fixed angular scale we see similar trends, with timescales increasing slightly at higher redshifts as the angular diameter distance begins to decrease for $z\gtrsim1.5$. The values in the plot correspond well with typical values for asymmetry timescales from idealised simulations \citep{Lotz2008,Lotz2010}, which range from a few tens of Myr up to around 300~Myr. 

\subsubsection{Dependence on the gas properties of the merger}
\label{sec:timsecales:satellites}

In this section, we consider the role of the physical properties of the gas of the merging pair of galaxies in determining the merger duration. In order to determine the mass of cold, potentially star-forming gas in the primary and secondary galaxy, we extract values of gas density and temperature from the \textsc{NewHorizon} AMR grid up to the maximum possible refinement ($\sim34$~pc in proper units). Gas properties are extracted in a spherical volume around the centre of both the secondary and primary galaxy identified in the merger trees at $t_{\rm max}$ and the gas distribution is extracted extending out to 3~R$_{\rm eff}$ from the centre of each galaxy. Since winds, ram-pressure and the environment have the potential to either trigger or quench star formation in any relatively cool gas \citep[e.g.][]{Scodeggio1993,Chung2009,Poggianti2016,Lee2017,Sheen2017}, we select relatively liberal criteria for defining the star-forming gas reservoir of a galaxy, requiring that a gas cell has a minimum density of 5~H$\,$cm$^{-3}$ and a maximum temperature of $10^{5}$~K (although tightening or further relaxing these criteria does not qualitatively change any of our results). We calculate gas fractions for each galaxy as follows:

\begin{equation}
    f_{\rm gas} = \frac{M_{\rm gas}}{M_{\rm tot}},
\end{equation}
\noindent where $M_{\rm gas}$ is the total gas mass within 3~R$_{\rm eff}$ and $M_{\rm tot}$ is the total baryonic mass within 3~R$_{\rm eff}$ (the total stellar and gas mass).

In Fig \ref{fig:timescales:gas}, we study merger durations as a function of the properties of the merging pair. The left panel is a scatter plot showing merger durations as a function of redshift (the same points that are used to calculate merger durations in Fig \ref{fig:timescales}). Points are colour coded by the gas fraction of the merging pair as described above and the size of each point corresponds to the stellar mass of the primary. Open circles indicate minor mergers and filled circles indicate major mergers. The inset plot shows the average gas fraction of the secondary (dashed) and primary galaxy (dotted) as a function of redshift. In order to investigate the dependence of merger duration on the gas properties of the pair, we select a sample of galaxies with very low gas fraction ($f_{\rm gas}<0.1$, red solid line). By controlling for $f_{\rm gas}$ in this way, we now only observe limited evolution in the merger duration as a function of redshift. By doing the same for mergers with high gas fractions (blue solid line) we observe longer merger durations than the average given by the solid grey line. 

As stars and gas are accreted from the secondary it takes time for this material to settle into a more ordered state within the primary galaxy. In the case of a gas poor merger, the merger ends once the accreted material has settled, whereas in the gas rich case, accreted gas can continue to form stars and drive continued disturbances in the primary even after the stellar component has settled. The increase in merger duration towards lower redshift in the high gas fraction sample is partially driven by the fact that the secondary galaxies become significantly more gas rich compared to the primary at low redshifts (see blue line in the inset plot). The large influx of gas can significantly enhance the \textit{relative} amount of star formation (typically for 250-500~Myr) and drive morphological disturbances in the primary when there is a more gas poor host compared to a merger between two galaxies that are equally gas rich. A second driver of the increase of merger durations towards low redshift is the increase in dynamical timescales, which are inversely proportional to density ($t_{dyn}\propto \rho^{-1/2}$). Since the background density scales as $(1+z)^3$, causing the galaxy virial radii to shrink towards higher redshifts and their average density to increase, galaxy dynamical timescales become significantly shorter. Finally at least a small part of this evolution is related to the fact that we define the morphological disturbance relative to the local standard deviation of the residual between the smoothed continuum fit of the axis length and the actual evolution of the axis length while the galaxy is not significantly disturbed (see Section \ref{sec:disturbances_def}). Since the morphologies of star forming galaxies are typically much more complicated and chaotic at high redshifts \citep[e.g][]{Abraham2001, Conselice2014}, $\sigma_{\rm local}$ is typically larger, meaning the threshold over which a galaxy is considered to not be disturbed is somewhat different at high redshift. In Appendix \ref{sec:test}, we explore in more detail how changes in this threshold change our results.

The right panel of Fig \ref{fig:timescales:gas} is similar to the left panel, but shows galaxy colours rather than gas fractions. Galaxy colours are a good proxy for gas content and therefore the duration of the merger remnant at lower redshifts. However, since galaxies can remain blue for some time after becoming gas poor, colour is a less reliable predictor of the merger duration at very high redshifts, where the time that it takes the blue stellar populations in quenched galaxies to fade is comparable to the age of the Universe \citep{Tinsley1980}. 

\section{Influence of mergers and interactions on instantaneous star formation rate}

\label{sec:starbursts}

In this section we study the instantaneous star formation rates of merging and interacting galaxies compared with the star forming main sequence (SFMS). Star formation rates are averaged over an interval of 100~Myr so that the star formation rate is given by ${\rm SFR} = \sum \{m_{\star} \mid a_{\star} < 100~{\rm Myr}\}/100~{\rm Myr}$, where $m_{\star}$ is the mass of each star particle and $a_{\star}$ is its age, varying the time interval between 10~Myr and 500~Myr does little to alter our results. We fit to the SFMS at each snapshot assuming that it is described by a power-law and excluding all quenched galaxies (i.e. galaxies with SFRs of 0). We then calculate the average displacement in star formation rates from the SFMS of the merging and disturbed galaxies. 

\begin{figure}
	\centering
	\includegraphics[width=0.45\textwidth]{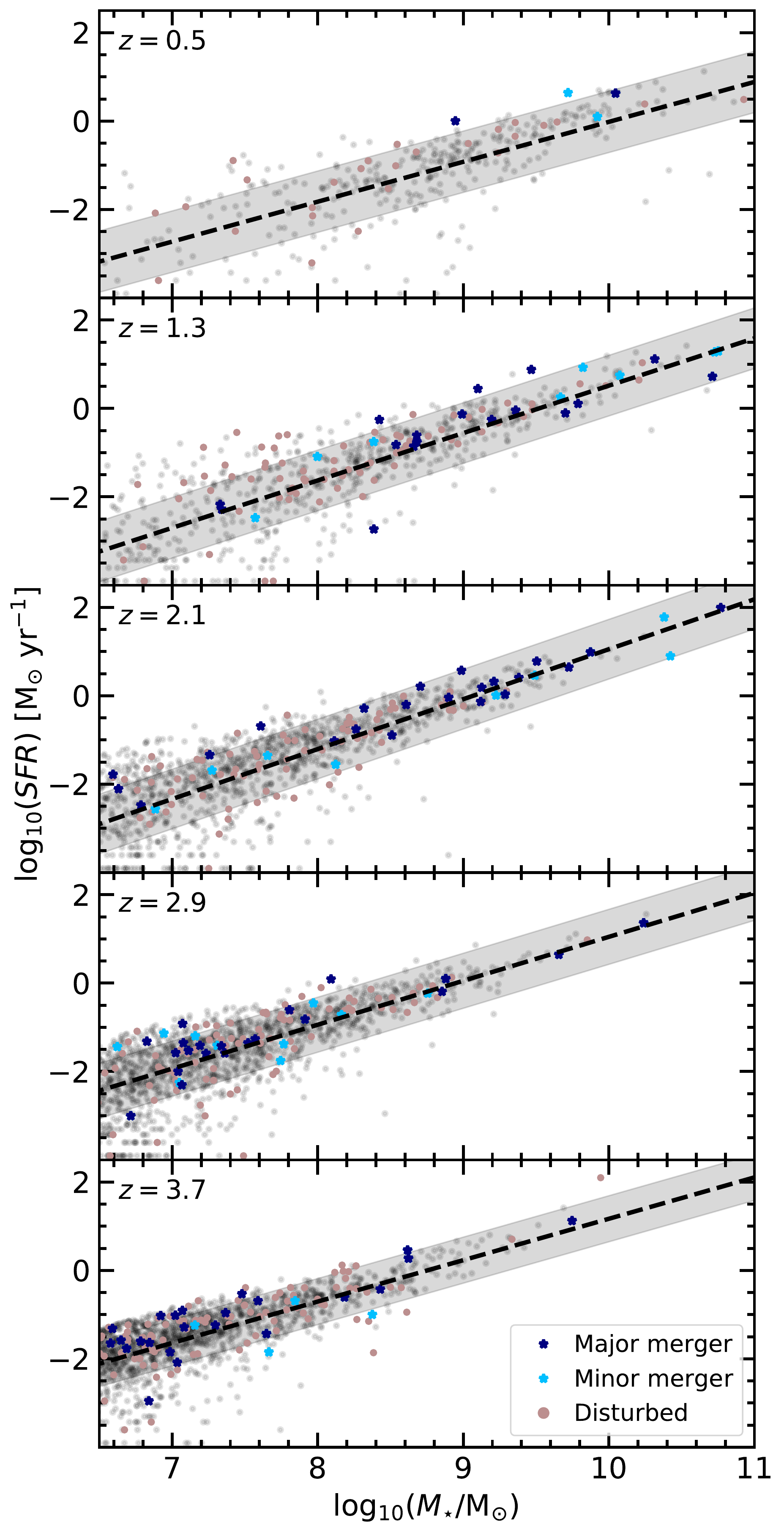}
    \caption{The star forming main sequence for different redshifts. Blue stars indicate galaxies that are merging and light red points indicate galaxies that are disturbed. Grey points are the remaining galaxies not classified in any of these sub-samples. The shaded region indicates the $1\sigma$ spread of the SFMS.}
    \label{fig:main_sequence}
\end{figure}

Fig \ref{fig:main_sequence} shows the evolution of the SFMS for snapshots from redshift 0.5 to 4. Dashed lines indicate the fit to the main sequence and dark blue stars, light blue stars and light red circles indicate galaxies that are undergoing major mergers, minor mergers or are morphologically disturbed as a result of other interactions respectively. In the low redshift universe ($z<2$), merging galaxies have clearly enhanced star formation on average, while disturbed galaxies in general are also visibly enhanced, they do not typically exhibit star formation rates quite as high above the main sequence as the most star forming mergers. Towards higher redshifts merging galaxies begin to fall onto the main sequence, with only moderate displacement above the main sequence after $z=2.5$. If we consider only galaxies in the 90th percentile of sSFR at snapshots where $z<1.5$, (redshifts where mergers produce the largest enhancements on average), we see that the most extreme merging and non-merging galaxies are both able to reach similarly high sSFRs, but merging galaxies are host to higher levels of star formation on average. This suggests that mergers usually act to only modestly enhance star formation in galaxies with relatively ordinary star formation rates, rather than drive very large enhancements that produce anomalously large sSFRs.

\begin{figure}
	\centering
    \includegraphics[width=0.45\textwidth]{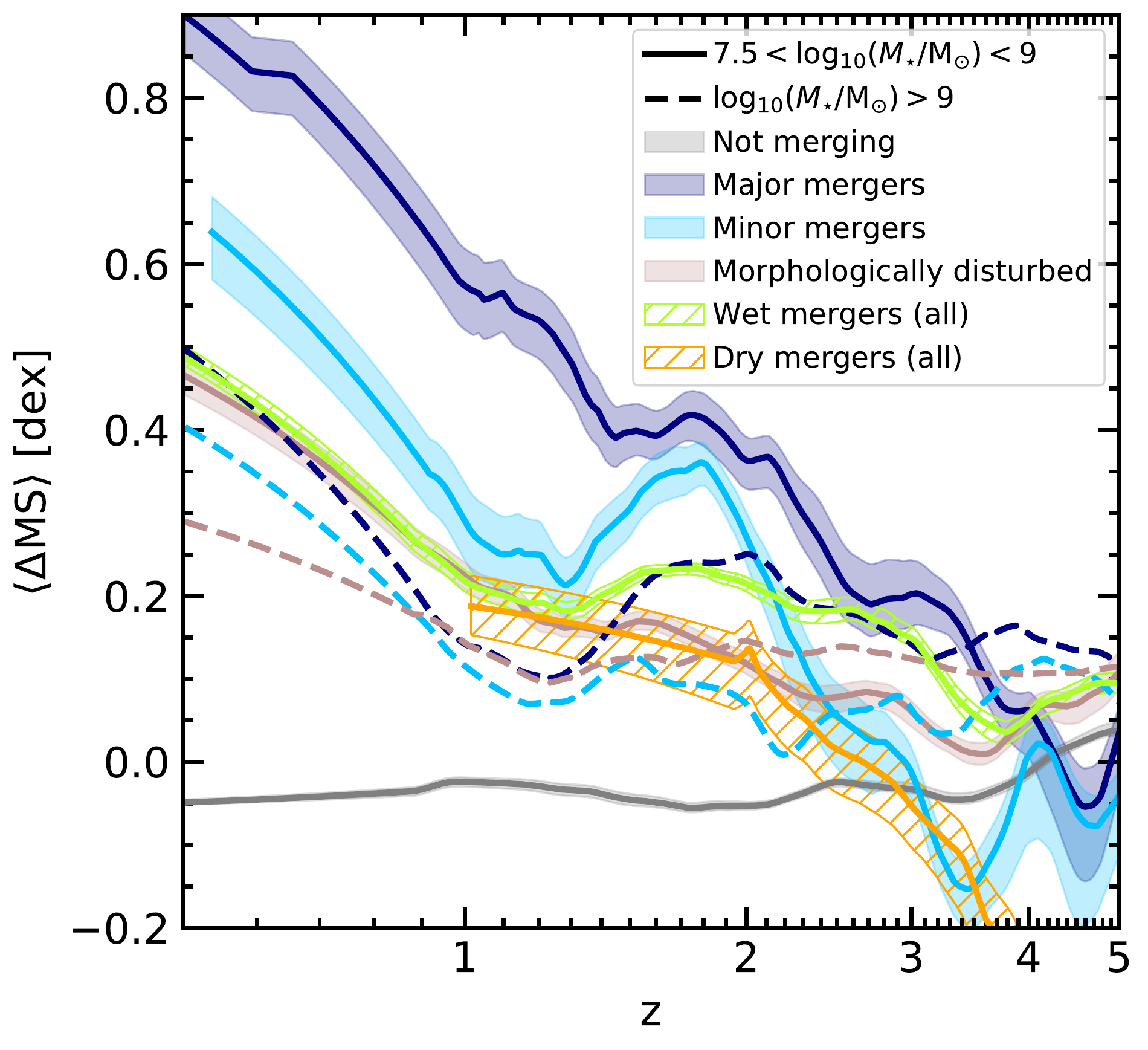}
    \caption{Average displacement from the main sequence in dex as a function of redshift. Dark blue lines indicate major mergers, light blue lines indicate minor mergers and light red points indicate morphologically disturbed galaxies. Solid lines correspond to dwarf masses ($7.5<{\rm log_{10}}(M_{\star}/{\rm M_{\odot}})<9$) with $1\sigma$ errors indicated by the shaded regions. For comparison, we also present the same for galaxies more massive than $10^{9}{\rm M_{\odot}}$. For clarity, errors are not shown but are similar to the dwarf sample. Green and orange lines and hatched regions show the average displacement for wet (combined $f_{\rm gas} > 0.25$) and dry mergers (combined $f_{\rm gas}<0.05$).}
    \label{fig:main_sequence_delta}
\end{figure}

Fig \ref{fig:main_sequence_delta} shows $\langle \Delta {\rm MS} \rangle$, the average displacement from the main sequence for merging and interacting galaxies in dex, as a function of redshift. As in Fig \ref{fig:main_sequence} dark blue, light blue and light red lines correspond to major and minor mergers and interactions respectively; solid lines indicate dwarf galaxies and dashed lines indicate more massive galaxies. In the case of both major and minor mergers and interacting galaxies, the star formation displacement increases higher above the main sequence towards lower redshifts. Major mergers are typically more star forming than minor mergers, while interacting galaxies have still more modest displacements from the main sequence (around 0.1 to 0.2 dex lower than minor mergers). Dwarf merging galaxies generally lie higher above the main sequence than their more massive counterparts, except towards high redshift ($z>3$), where mergers in dwarf galaxies produce, in some cases, a slight reduction in star formation compared to the main sequence. Interacting galaxies, on the other hand, continue to enhance star formation, at least to a small degree.

Green and orange lines with hatched regions indicate $\langle \Delta {\rm MS} \rangle$ for wet (combined $f_{\rm gas} > 0.25$) and dry mergers (combined $f_{\rm gas}<0.05$) respectively as measured just before $t_{\rm max}$. As expected \citep[e.g.][]{Bell2006,Lin2008}, dry mergers do very little to enhance star formation rates at any redshift (although dry mergers account for only 4 per cent of mergers over the redshift range that we consider). Wet mergers typically produce similar enhancements regardless of gas fraction. All mergers with gas fractions greater than 0.15 produce similarly modest enhancements on average.

The evolution of star formation enhancement over redshift that we see in the dwarf sample, corresponds well with the observed star formation enhancement of their high mass counterparts in the observed Universe. For example, mergers are able to drive orders-of-magnitude enhancements in SFRs of galaxies in the nearby Universe \citep[e.g.][]{Armus1987,Duc1997,Elbaz2003}, but observational and theoretical evidence suggests mergers between massive galaxies in the high redshift Universe produce only weak or negligible enhancement of star formation \citep[e.g.][]{Kaviraj2013,Lofthouse2017,Fensch2017,Martin2017}. The generally high star formation rates in these epochs are likely instead driven by high molecular gas fractions that are the result of intense cosmological gas accretion \citep[e.g.][]{Tacconi2010,Geach2011,Bethermin2015}.

\section{Dwarf galaxy assembly from mergers and interactions}

\label{sec:assembly}

In this section we consider the influence of various mechanisms on the mass assembly of dwarf galaxies ($M_{\star} < 10^{9}$~M$_{\odot}$), namely, direct accretion of stellar mass formed from ex-situ sources and enhancements of in-situ star formation triggered by non merger interactions, major and minor mergers.

\subsection{Direct assembly by merging}
\label{sec:direct_accretion}

Here we investigate the proportion of stellar mass growth in dwarf galaxies that is driven directly by \textit{ex-situ} accretion of stellar mass via mergers with lower mass galaxies. We use the merger catalogues described in Section \ref{sec:merger_catalogues} to obtain the stellar mass at $t_{\rm max}$ of each secondary galaxy that merges with one of the $z=\basez$ main progenitors. The total stellar mass evolution of each galaxy is obtained by following the chain of each main progenitor. We assume that all star formation not accounted for by merging (i.e. the net star formation after subtracting the contribution from mergers) is the result of stars formed \textit{in-situ} either as a result of secular star formation or formed during merger or interaction driven episodes.

We first note that, since the finite mass resolution of the simulation makes it impossible to consider mass accreted from mergers of all mass ratios, we instead only consider mergers with mass ratios greater than 1:10. As Fig \ref{fig:completeness} shows, such mergers are generally resolved by the simulation across the full range of masses and redshifts that we consider ($M_{\star}>10^{7.5}$~M$_{\odot}$, $0.5<z<5$). We check how the inclusion of lower mass ratio mergers may affect our results by considering a sample of galaxies with $z=\basez$ main progenitor masses between $8<{\rm log_{10}}(M_{\star}/{\rm M_{\odot}})<9$ for redshifts of $z<2$. The stellar masses of the galaxies in this sample remain more than 100 times the minimum detectable object mass of $4\times10^{5}$~$M_{\odot}$ over the whole redshift range we consider. We find that 60 per cent of accreted stellar mass is explained by $R>1:10$ mergers alone, leaving around $\sfrac{2}{5}$ of stellar mass unaccounted for. The fraction of missing mass is likely to be less severe for the rest of the dwarf galaxy sample, since the median merger mass ratio moves closer to 1:1 as we move to very low stellar masses. However, if we assume a value of $\sfrac{2}{5}$ for the whole sample, this missing mass does little to alter our conclusions about the significance of ex-situ stellar mass accretion in the dwarf regime.

\begin{figure}
	\centering
    \includegraphics[width=0.45\textwidth]{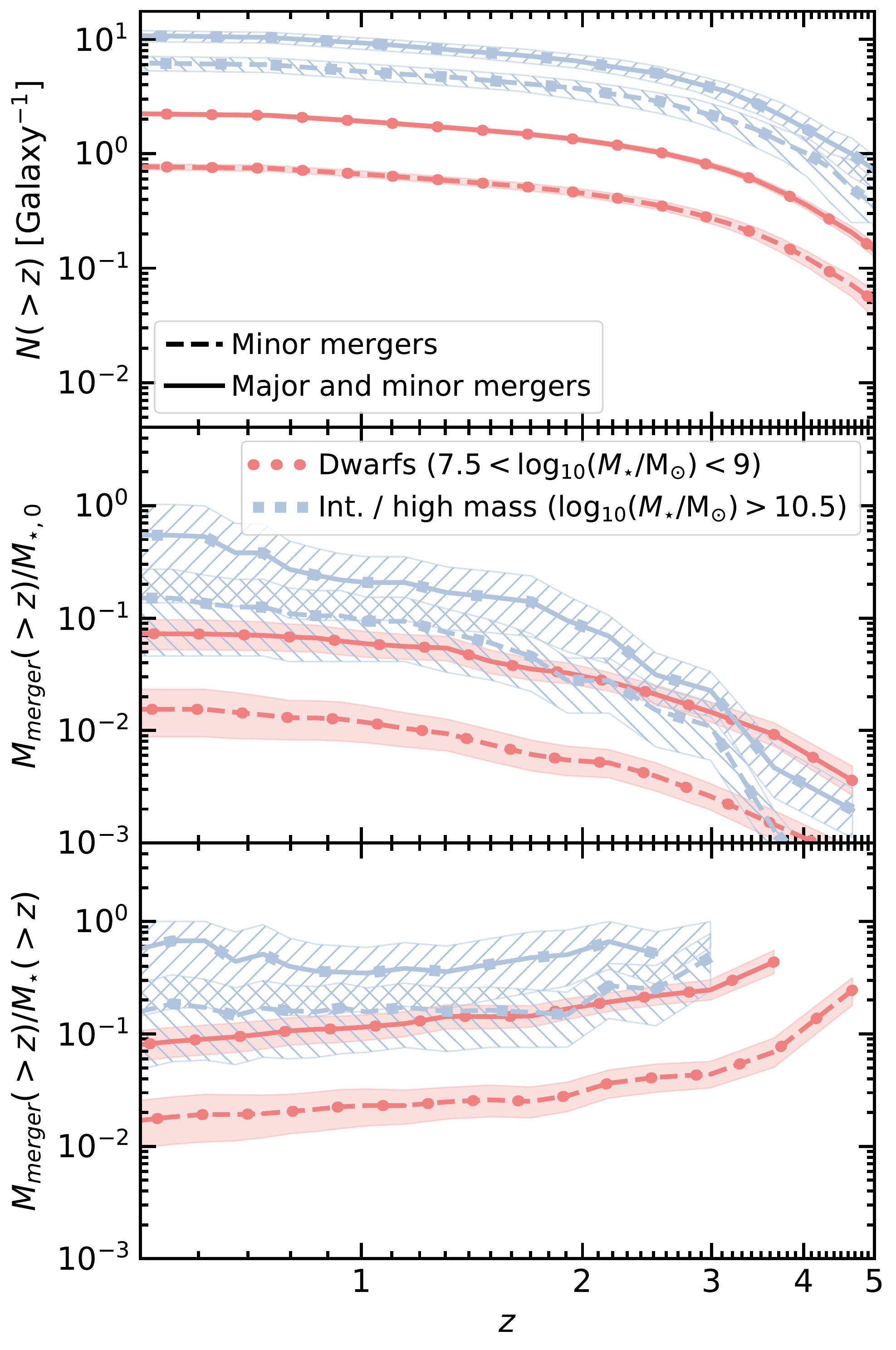}
    \caption{\textbf{Top}: cumulative number of minor (dashed line) and total number of mergers with mass ratios greater than $1:10$ per galaxy for dwarfs (light red lines) and intermediate mass galaxies (light blue lines). \textbf{Middle}: cumulative mass accreted from minor and $1:10$ or greater mass ratio mergers as a fraction of the total stellar mass at $z=\basez$. \textbf{Bottom}: ratio of the cumulative accreted mass and cumulative total stellar mass. Shaded regions and hatched regions show the $1\sigma$ uncertainties.}
    \label{fig:merger_accretion}
\end{figure}

Fig \ref{fig:merger_accretion} shows the merger rates and stellar mass accreted from mergers for a sample of dwarf galaxies ($7.5<{\rm log_{10}}(M_{\star}/{\rm M_{\odot}})<9$; red lines), consisting of 648 galaxies. For comparison, we also study a sample of intermediate and high mass galaxies (${\rm log_{10}}(M_{\star}/{\rm M_{\odot}})>10.5$; blue lines). The intermediate/high mass sample consists of 15 galaxies the most massive of which has $M_{\star} = 10^{11.2}~{\rm M_{\odot}}$.

The top panel shows the cumulative number of mergers undergone per galaxy for minor mergers (dashed lines) and all mergers with $R>1:10$ (solid lines). In the dwarf regime, mergers are quite rare, with each galaxy undergoing only one or two $R>1:10$ mergers in their lifetime on average (typically one major and one minor) compared with the intermediate/high mass galaxies which undergo an order of magnitude more. This is also compatible with the work of \citet{Deason2014}, who show that only 10 per cent of dwarf galaxies have undergone mergers with mass ratios greater than 1:10 between $z=1$ and the present day.

The middle panel shows the cumulative merger accreted stellar mass as a fraction of the total stellar mass of the sample's $z=\basez$ progenitors. Only 7.5 per cent (or 10 per cent assuming $\sfrac{2}{5}$ of mass comes from lower mass ratio mini mergers) of the stellar mass comes directly from mergers in the dwarf regime, mirroring the results of \citet{Fitts2018}, who also find that less than 10 per cent of dwarf galaxy stellar mass is formed ex-situ. On the other hand, in the intermediate/high mass regime, around 40 per cent of stellar mass is formed ex-situ, similar to values found in other work \citep[e.g.][]{Oser2010,Dubois2013,JLee2015,Rodriguez2016,Dubois2016,Tacchella2019,Davison2020}. Major and minor mergers bring in roughly equal amounts of stellar mass, although major mergers are somewhat more important in the dwarf regime than at intermediate/high mass.

Finally, the bottom panel shows the ratio of the cumulative ex-situ stellar mass and the cumulative total stellar mass for each sample. As the downward trend of the black lines indicates, mergers become far less important to the mass assembly of galaxies in the dwarf regime over time, with mergers dominating their mass assembly only at very high redshifts. Although the rate of at which ex-situ mass is accreted remains roughly constant over time ($\dot M / M_{z=\basez}\approx0.15$~Gyr$^{-1}$), their star formation rate accelerates as they increase in mass through secular accretion of cold gas, quickly overtaking the contribution from mergers. Thus, while mergers may be important drivers of dwarf galaxy evolution very early on when the merger rate is high, this contribution is quickly washed out, especially as we are considering only galaxies that survive to $z=\basez$. Although a similar evolution is seen initially in the intermediate/high mass sample, this eventually flattens as their average star formation rate slows. Therefore, mergers remain important in the higher mass regime, although their contribution is expected to become increasingly important at higher masses than those probed here, particularly above stellar masses of $10^{11}$M$_{\odot}$ \citep{Dubois2016,Rodriguez2016,Davison2020}.

\subsection{Assembly by merger and interaction induced star formation}
\label{sec:induced_star_formation}

While we have shown that galaxy mergers do not drive a particularly significant proportion of stellar mass growth through direct accretion in the dwarf regime, it is still worth considering how mergers and other interactions might influence their stellar mass growth less directly. In this section, we investigate how interactions and mergers might drive the stellar mass growth of dwarf galaxies through induced star formation.

\subsubsection{Merger and interaction driven enhancement of star formation}
\label{sec:merger_enhancement}

We first define the star formation enhancement of the merging population, $\xi$, at each snapshot as the ratio of the mean sSFR of the merging population and the population of galaxies that is not merging or disturbed at a given snapshot. As in Section \ref{sec:timsecales}, galaxies are considered merging using the definition in Section \ref{sec:disturbances_def} -- mergers begin at $t_{\rm max}$ and end following $\tau_{\rm merger}$. In order to account for the dependence of the enhancement as well as the weak dependence of sSFR with stellar mass \citep[e.g.][]{Whitaker2012}, $\xi$ is measured in bins of both redshift and stellar mass:

\begin{equation}
\label{eqn:xi}
	\xi(M_{\star},z) = \frac{\big\langle \mathrm{sSFR}_{\rm m}(M_{\star},z) \big\rangle}{\big\langle \mathrm{sSFR}_{\rm nd}(M_{\star},z) \big\rangle},
\end{equation}

\noindent where $\big\langle \mathrm{sSFR}_{\rm m}(M_{\star},z) \big\rangle$ is the average sSFR of the merging population in a given mass and redshift bin (the same mass bins as those shown in Fig \ref{fig:completeness}). $\big\langle \mathrm{sSFR}_{\rm nd}(M_{\star},z) \big\rangle$ is calculated in the same way for the population of galaxies that is not disturbed or merging.

Since $\xi(M_{\star},z)$ tells us the specific star formation rate of the merging population relative to a control population of non-merging galaxies, which we can use to estimate the fraction of star formation that is driven \textit{directly} by mergers. The excess stellar mass formed in the merging population is given by:

\begin{equation}
    \delta(M_{\star},z) = m_{\rm new,m}(M_{\star},z) \big[1- 1/ \xi(M_{\star},z)\big],
\end{equation}

\noindent where $m_{new,m}(M_{\star},z)$ is the total stellar mass formed in merging galaxies in each stellar mass and redshift bin. The fraction of star formation attributable to mergers in a given mass and redshift bin is therefore given by	$\delta(M_{\star},z)/m_{\rm new}(M_{\star},z)$, where $m_{{\rm new}}(M_{\star},z)$ is the total stellar mass formed in galaxies regardless of whether they are merging in each of the stellar mass and redshift bin in question within the same time interval. We note that the results presented in this section are robust to the merger timescale used. By choosing fixed merger timescales of 0.5~Gyr and 1~Gyr, we do not see a significant change in the overall merger driven star formation budget (see Appendix \ref{sec:test}).

The \textit{interaction driven} star formation enhancement is calculated almost identically to merger driven star formation enhancement, except that the numerator of Eqn \eqref{eqn:xi} includes all morphologicaly disturbed galaxies, which are not also merging. We assume that morphological disturbances that are not the result of mergers must instead be due to some other type of non-merger interaction.

\subsubsection{Average enhancement and contribution to star formation}

\begin{figure}
	\centering
    \includegraphics[width=0.45\textwidth]{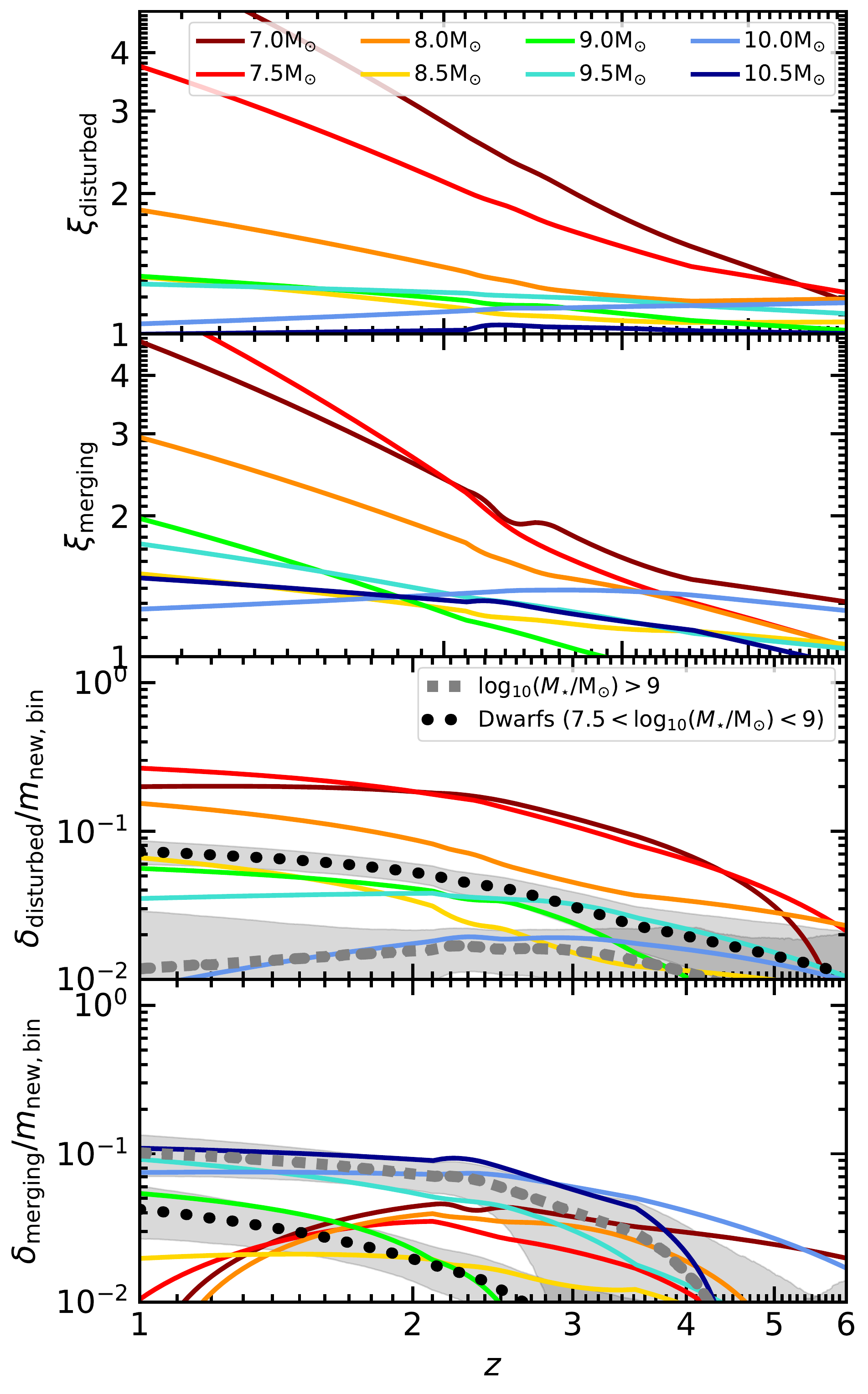}
    \caption{The first and second panels show the average fractional enhancement in sSFR for disturbed (interacting) galaxies and merging galaxies compared to the non-disturbed population in bins of progenitor stellar mass. The third and fourth panels show the fractional SFR excess due to interactions and mergers respectively. Coloured lines indicate the evolution of each quantity with redshift for different stellar masses and lines dotted with black circles and grey squares show the average excess due to interactions or mergers for dwarf masses and $M_{\star}>10^{9}$~M$_{\odot}$ respectively weighted by the stellar mass formed in each bin.}
    \label{fig:average_enhancement}
\end{figure}

Fig \ref{fig:average_enhancement} shows the evolution of galaxy star formation enhancement and the average fraction of stellar mass formed as a result of mergers or interactions as a function of redshift. The first and second panels show the average enhancement in specific star formation rate for merging and morphologically disturbed populations. The average star formation enhancement produced as a result of interactions is relatively modest while mergers produce significantly larger enhancements. In both cases lower mass galaxies exhibit significantly larger enhancements than high mass galaxies, particularly at lower redshifts. These results appear consistent with observational studies by \citet{Stierwalt2015}, which show average star formation enhancements of $2.3\pm0.7$ times in isolated dwarf pairs with separations of $<50$~kpc. If we combine mergers and interactions in the \textsc{NewHorizon} sumulation, weighting our results according to the mass distribution shown in \citet[][Fig 1]{Stierwalt2015}, we obtain a similar value for $\epsilon$ in the dwarf regime of 2.2. We stress that these measurements are not exactly analogous as our redshift range and definition of what constitutes a merger or interaction differ. By varying the threshold for morphological disturbance between to 1.5~$\sigma_{\rm mult}$ and 3.5~$\sigma_{\rm mult}$, we obtain average enhancements varying between 1.8 and 2.6.

The third panel shows the excess SFR of the morphologically disturbed galaxies divided by average SFR of all galaxies in a given bin as a function of redshift and stellar mass. While SFR enhancement is relatively modest (as Fig \ref{fig:morphological_disturbance} shows), galaxies of all masses spend a significant fraction of their lifetime in a morphologically disturbed state. Because dwarf mass galaxies have larger SFR enhancements but spend roughly the same amount of time as more massive galaxies in a morphologically disturbed state, significantly more star formation is triggered by interactions than their more massive counterparts.

Finally, the bottom panel shows the average excess SFR of the merging galaxies divided by average SFR of all galaxies as a function of redshift and stellar mass. Although mergers produce very significant enhancements in the star formation rates of low mass galaxies (especially $10^{7.5}$~M$_{\odot}$, where there are enhancements of 3-4 times at $z=1$), this is not enough to produce a significant increase in the average merger driven star formation budget. This can be explained by the fact the dwarf-dwarf mergers are comparatively rare (recall that as Section \ref{sec:direct_accretion} shows, the merger rate for galaxies more massive than $10^{10.5}$~M$_{\odot}$ is around an order of magnitude higher than the dwarf population), meaning the net contribution from mergers remains small.

\begin{figure}
	\centering
    \includegraphics[width=0.45\textwidth]{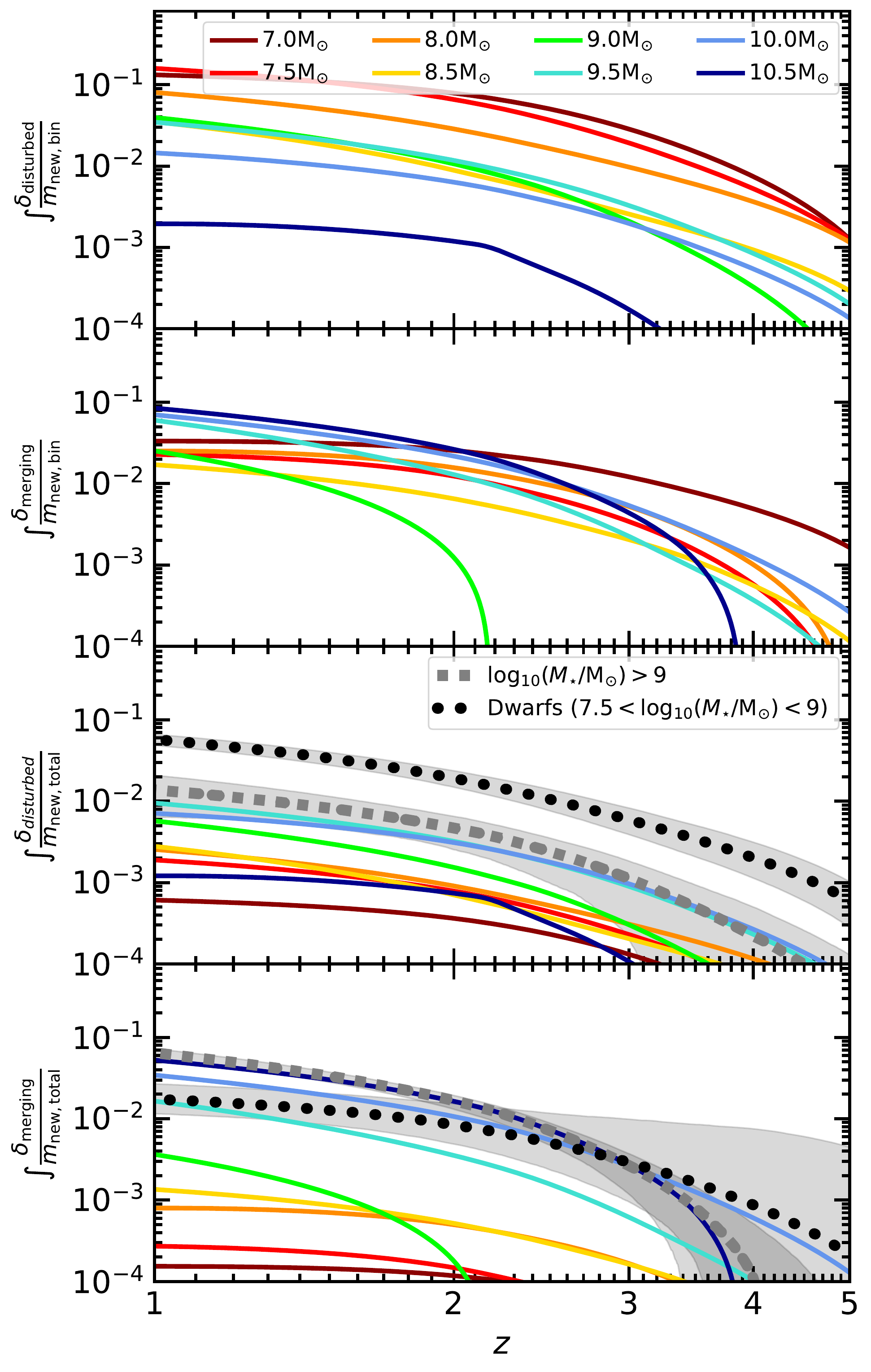}
    \caption{The first and second panels show the cumulative SFR budget that is due to non-merger interactions or mergers. The third and fourth panels show the fractional SFR excess due to interactions and mergers as a fraction of the total star formation \textit{within a given mass bin}. The third and fourth panels show the same but as a fraction of the  \textit{total} star formation budget. Coloured lines indicate the evolution of each quantity with redshift for different stellar masses and lines dotted with black circles and grey squares show the average contribution to the star formation budget due to interactions or mergers for dwarf masses and $M_{\star}>10^{9}$~M$_{\odot}$ respectively.}
    \label{fig:budget}
\end{figure}

Fig \ref{fig:budget} shows the cumulative fraction of the star formation budget driven by interactions and mergers as a fraction of the stellar mass formed only in the given mass bin (first and second panel) and the cumulative star formation budget driven by interactions and mergers in each mass bin as a fraction of the total star formation budget over all mass bins (third and fourth panel). Within their respective mass bins, interactions can drive a significant fraction of star formation in dwarf galaxies. Mergers, on the other hand, drive a larger fraction of the star formation budget for the higher mass galaxies. In total, interactions drive 7 per cent of stellar mass growth for dwarf galaxies while mergers drive only 2 per cent. The reverse is true when we look only at massive galaxies, where mergers drive 8 per cent of star formation and interactions drive only 1 or 2 per cent. In Appendix \ref{sec:test}, Fig \ref{fig:budget_fixed} we show the bottom panel of Fig \ref{fig:budget} using a fixed merger timescale rather than timescales calculated in Section \ref{sec:disturbances_def}. Although their evolution differs, the final budget at $z=1$ of 6 per cent / 9 per cent for 0.5~Gyrs and 6 / 10 per cent for 1~Gyr for the dwarf and $M_{\star}>10^{9}$~M$_{\odot}$ samples do not differ significantly from our results here and do not qualitatively change our results.

The results we find for higher mass galaxies are broadly consistent with other recent work \citep[e.g.][]{Robaina2009,Lamastra2013,Fensch2017,Martin2017,Patton2020} which indicate that the sSFR enhancement and contribution of mergers to the cosmic star formation budget is modest (sSFR enhancements up to a few times the average and around 10 per cent of the star formation budget at $z<2$), except at low redshifts where star formation has slowed significantly since the peak of the cosmic star formation rate density \citep[e.g.][]{Madau2014}.

In this section we have explored how mergers and interactions drive the assembly of dwarf galaxies over cosmic time, we found that no single process is responsible for producing more than 10 per cent of the total stellar mass in this mass range. However, when we consider the contributions from all sources together, we find that the aggregate contribution in the dwarf regime is non-negligible. We briefly summarise our findings below.

In the dwarf regime, ex-situ mass from major and minor mergers accounts for around 10 per cent of the total mass of each galaxy. Major and minor mergers also drive an additional 2 per cent through star formation enhancements induced in the merger and post merger phases. Finally, non-merger interactions like fly-bys are responsible for an additional 7 per cent of their mass through star formation enhancements. In total around 20 per cent of the stellar mass in dwarf galaxies at $z=\basez$ was formed as a result of these processes, with the remaining stellar mass forming as a result of secular processes.

\section{Summary}

\label{sec:summary}

In this paper we have investigated the role of interactions and mergers in driving the morphology, star formation and mass assembly of dwarf galaxies using the \textsc{NewHorizon} cosmological hydrodynamical simulation. We have also performed a full accounting of the contribution of mergers and other interactions to the assembly of dwarf galaxies. Our key results are as follows:

\begin{enumerate}
\item \textit{Galaxies of all stellar masses spend a significant proportion of their lifetime in an unrelaxed or morphologically disturbed state.} On average, interactions (mergers of fly-bys) drive significant variations in the axis lengths of galaxies which result in them being in an unrelaxed state for 10 per cent to 30 per cent of a galaxy's total lifetime a result. The fraction of morphologically disturbed galaxies falls gently towards lower redshifts, with 5 per cent of dwarfs having disturbed morphologies at $z=1$ compared to around 20 per cent at $z=3$.
\\
\item \textit{Mergers drive a very limited amount of the morphological disturbances seen in dwarf galaxies.} The vast majority of morphological disturbances in dwarf galaxies are driven by non-merger interactions. At no redshift after $z=5$ are mergers responsible for a larger proportion of morphological disturbances than non-mergers. Compared to more massive galaxies ($M_{\star}>10^{10.5}$~M$_{\odot}$), for which close to 100 per cent of morphological disturbances are the result of major or minor mergers, less than 10 per cent of morphological disturbances in dwarf galaxies are driven by mergers. This finding is particularly important to future studies that may attempt to identify dwarf mass mergers and post mergers morphologically using instruments like JWST given the large fraction of apparently morphologically disturbed dwarf galaxies which are not, in fact, merging. 
\\
\item \textit{The time that merger remnants take to relax increases towards low redshift.} This is partially driven by the fact that the lower mass secondary galaxies remain relatively gas rich compared to the primary at low redshifts, driving larger relative increases in star formation as well as the fact that dynamical timescales are significantly shorter at high redshifts. At high redshift, merger durations are slightly influenced by the fact that our baseline for when a merger remnant can be considered to be relaxed evolves slightly with redshift due to the generally chaotic nature of galaxy kinematics and morphology at high redshifts.
\\
\item \textit{Non parametric measures like $Gini$-$M_{20}$ and $A_{g}$-$C$ are not effective at separating mergers from otherwise morphologically disturbed objects.} While major mergers, minor mergers and disturbed galaxies can be disentangled from the undisturbed population, they occupy similar regions of parameter space for both $A_{g}$-$C$ and $Gini$-$M_{20}$. Since neither of these methods are sensitive to merger relics like tidal tails, which carry only a few per cent of the total stellar mass of the galaxy, it may still be possible to differentiate between mergers, fly-bys and other interactions through the presence of tidal features given deep enough imaging.
\\
\item \textit{Mergers and interactions in the dwarf regime typically drive moderate increases in the star formation rate for low and intermediate redshifts only.} Towards higher redshifts ($z>3$), mergers do not have any appreciable impact on the star formation rate when compared to the non-merging population. Non-merger interactions drive small enhancements in the star formation rate, however, these types or interactions are also much more numerous than mergers in the dwarf regime.
\\
\item \textit{On average dwarf galaxies undergo one major merger and one minor merger on average between $z=5$ and $z=\basez$}. For comparison, for higher mass galaxies in the simulation ($10^{10.5}<M_{\star}/{\rm M}_{\odot}<10^{11.2}$) the average number of mergers with mass ratios greater than 1:10 is around 10. Mass assembly from the accretion of stellar mass formed ex-situ accounts for around 10 per cent of total dwarf stellar masses compared with 40 per cent for the higher mass sample.
\\
\item \textit{In-situ star formation enhancements driven by mergers and interactions account for another 10 per cent of the mass of dwarf galaxies}. Together mergers and interactions drive an additional 10 per cent of star formation, with interactions driving the vast majority. In aggregate, 20 per cent of the stellar mass in dwarf galaxies at $z=\basez$ was formed as a result of either mergers or other interactions, with the remaining 80 per cent of stellar mass forming as a result of secular processes.
\end{enumerate}


\section*{Acknowledgements}

We thank Clotilde Laigle for maintaining a version of the \textsc{sunset} code originally developed by Roman Teyssier. GM thanks Dennis Zaritsky, Richard Donnerstein and Gurtina Besla for interesting and fruitful discussions. This work has made use of the Horizon cluster on which the simulation was post-processed, hosted by the Institut d'Astrophysique de Paris. We warmly thank S.~Rouberol for running it smoothly. This work was granted access to the HPC resources of CINES under the allocations 2013047012, 2014047012, 2015047012, c2016047637 and A0020407637 made by GENCI and KSC-2017-G2-0003 by KISTI, as a `Grand Challenge' project granted by GENCI on the AMD-Rome extension of the Joliot Curie supercomputer at TGCC and under the allocation 2019-A0070402192 made by GENCI. Additional HPC resources were also kindly provided by KISTI (KSC-2017-G2-003), and large data transfer was supported by KREONET, which is managed and operated by KISTI. RAJ and SK acknowledge support from STFC grants ST/R504786/1 and ST/S00615X/1. SK acknowledges a Senior Research Fellowship from Worcester College Oxford. JD acknowledges funding support from the Beecroft Trust and STFC. TK was supported in part by the National Research Foundation of Korea (NRF2017R1A5A1070354 and NRF-2020R1C1C100707911) and in part by the Yonsei University Future-leading Research Initiative (RMS2-2019-22-0216). SKY acknowledges support from the Korean National Research Foundation (NRF-2020R1A2C3003769). This research is part of the Segal ANR-19-CE31-0017 (\href{https://www.secular-evolution.org}{https://www.secular-evolution.org}) and Horizon-UK projects. We thank the anonymous referee whose comments helped improve the clarity of this paper.

\section*{Data Availabilty}

The simulation data analysed in this paper were provided by the \textsc{NewHorizon} collaboration. The data will be shared on request to the corresponding author, with the permission of the \textsc{NewHorizon} collaboration or may be requested from \href{https://new.horizon-simulation.org/data.html}{https://new.horizon-simulation.org/data.html}.




\bibliographystyle{mnras}
\bibliography{paper_mnras} 

\appendix

\section{The primary driver of morphological disturbances}

\subsection{Merger tree convergence with Horizon-AGN}
\label{convergence}

\begin{figure}
	\centering
    \includegraphics[width=0.45\textwidth]{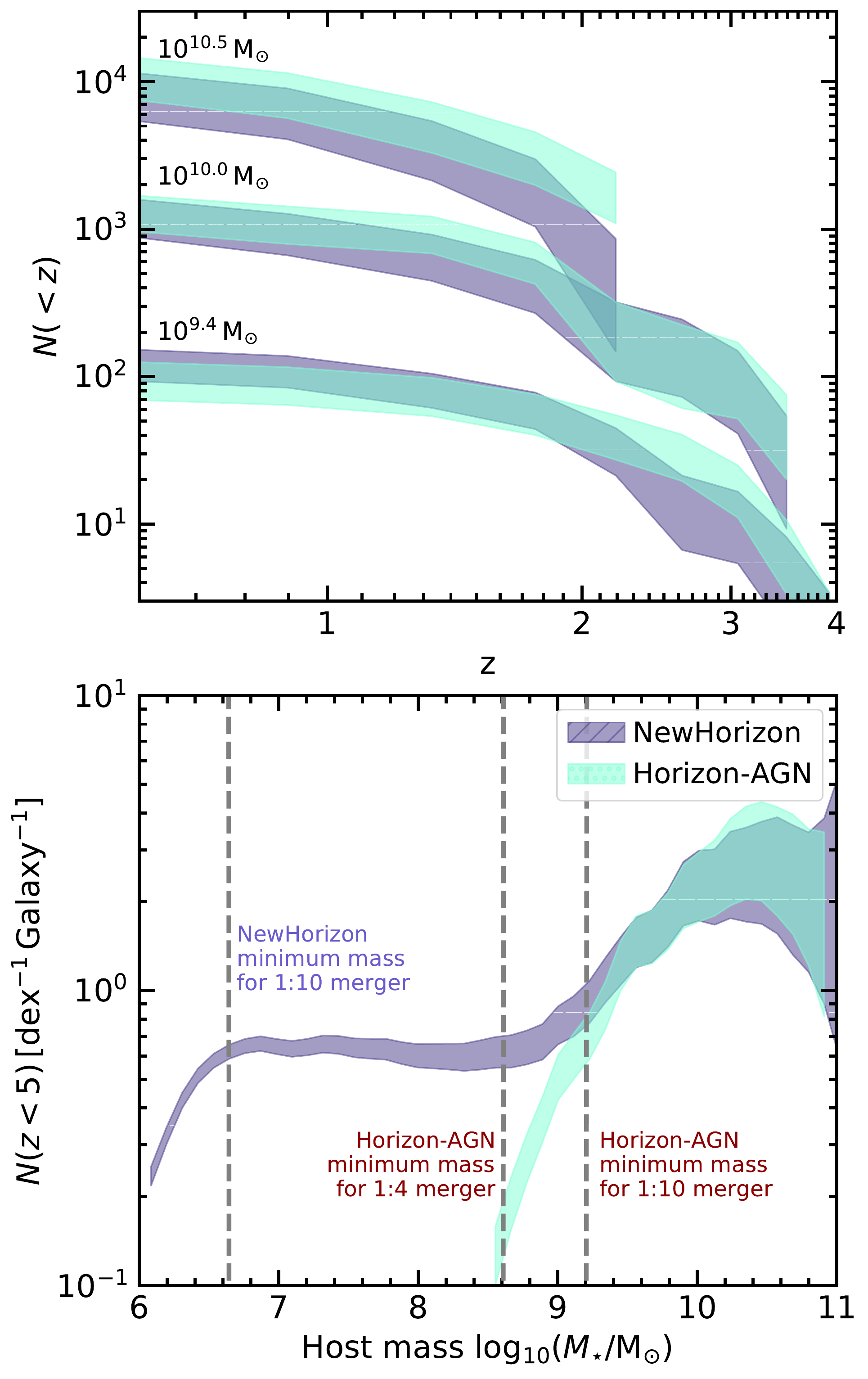}
    \caption{\textbf{Top}: Cumulative number of mergers occurring as a function for bins of stellar mass for the \textsc{NewHorizon} (blue) and Horizon-AGN (aquamarine) simulations. Shaded regions show the $1\sigma$ confidence interval for the cumulative number of mergers as a function of redshift and the centre of each 0.5 dex wide stellar mass bin is indicated above each region. Values on the $y$-axis are arbitrary and each bin is separated by an arbitrary amount \textbf{Bottom}: Total number of mergers undergone since $z=5$ as a function of stellar mass of the primary at the time of the merger and normalised by the total number of galaxies in the stellar mass bin.}
    \label{fig:merger_test}
\end{figure}

We first confirm that significant numbers of mergers are not missed and are therefore not misattributed to other processes. In order to determine whether mergers are consistently detected close to the resolution limit, we compare the merger rate (mass ratio $< 1:10$) for galaxies with a minimum stellar mass of $10^{9.2}~{\rm M_{\odot}}$ up to $\sim10^{11}~{\rm M_{\odot}}$ in the \textsc{NewHorizon} simulation with the merger rate of galaxies in the Horizon-AGN region matching the 10~Mpc spherical zoom-in region of \textsc{NewHorizon}. $10^{9.2}~{\rm M_{\odot}}$ is 10 times the minimum detectable mass in the Horizon-AGN simulation and therefore the limit at which any minor mergers are detectable, but 4 orders of magnitude larger than the minimum detectable structure mass in the \textsc{NewHorizon} simulation. Fig \ref{fig:merger_test} shows a comparison of merger rates in the \textsc{NewHorizon} and Horizon-AGN simulations. The first panel shows the cumulative number of mergers in several mass bins down to the minor merger limit of the Horizon-AGN simulation. The cumulative number of mergers remain consistent within the $1\sigma$ confidence intervals at all redshifts. Some degree of variation is to be expected, since a number of factors beyond resolution can lead to differences in merger rates. These include under-resolved galaxy formation in lower density regions leading to a deficit in or late formation of low-mass galaxies \citep[see][Fig 1]{Chabanier2020}, inherent stochasticity in galaxy properties and merger rates that arise from numerics or small differences in initial conditions \citep{Rey2018,Genel2019} and differences in sub-grid recipes that can lead to differences in galaxy properties and their evolution over cosmic time. 

The second panel shows the total number of mergers that have occurred since $z=5$ as a function of the stellar mass of the primary and normalised by the number of galaxies in each stellar mass bin. The lines for the \textsc{NewHorizon} and Horizon-AGN simulations are in good agreement up to the point where the number of detected mergers in the Horizon-AGN simulation are expected to decline due to the minimum mass imposed by the structure finder ($10^{8.2}~{\rm M_{\odot}}$). Similarly, the number of mergers with mass ratios greater than 1:10 detected in \textsc{NewHorizon} starts to decline for a primary stellar mass of  $4\times 10^{6}$~M$_{\odot}$ (corresponding to a 1:10 merger with a secondary mass of  $4\times 10^{5}$~M$_{\odot}$, which is the structure finder  minimum mass). \textsc{NewHorizon} therefore shows good convergence with Horizon-AGN both in terms of the global merger rate as a function of redshift and galaxy merger history as a function of stellar mass for the mass ranges where the two simulations overlap.

\subsubsection{Unresolved stellar components or dark satellites}

We also consider morphological disturbances that could arise from mergers with dark satellites \citep[e.g.][]{Helmi2012,Starkenburg2015} or haloes with stellar masses that would fall below the resolution limit, since these types of mergers would be absent from the galaxy merger trees but might still impart some some degree of change in galaxy morphology. We first select haloes with around one tenth the mass of the average halos that host the galaxies in the lowest stellar mass bin mass bin ($M_{\star}\sim10^{8}$~M$_{\odot}$) at $z=1$ (i.e. haloes massive enough to conceivably produce significant morpological disturbances if they were to merge). Within these haloes, we measure the total stellar mass within a virial radius, finding that only a small minority (4 per cent) of these haloes contain stellar masses below the limit of $4\times10^{5}$~M$_{\odot}$ (i.e. the stellar component of the vast majority of these haloes is detected). Mergers where the halo mass of the secondary would be likely to produce some visible disturbance but which contain no or very little stellar mass are therefore quite rare in this regime.

\subsection{The processes driving morphological disturbances}
\label{sec:PI_test}

In order to determine which processes are responsible for driving morphological disturbances, we first visually inspect the stellar particle distribution in a 200~kpc$\,\times\,$200~kpc box centred around 100 random disturbed galaxies over a 500~Myr window. In the majority of cases (90 per cent) we are able to confirm that the galaxy was morphologically disturbed as a result of an interaction with another galaxy at some point within this time frame. In most cases, a nearby companion of comparable or smaller mass that is responsible for producing the morphological disturbance through a fly-by was identified. The remaining cases are instead the result of strong tidal fields, tidal debris from an ongoing merger which merge later than the main body of the secondary or the misidentification of objects with no visible global disturbances (the result of intense, clumpy star formation or a poor continuum fit).

We also compare the PI, described in Eqn \eqref{eqn:PI}, of undisturbed and disturbed but non-merging galaxies. At each snapshot, we identify the galaxies that are disturbed as a result of non-merger processes and select a control sample of undisturbed galaxies with a matching stellar mass distribution. For galaxies in both samples we take a 200~Myr window and measure (1) the total integrated PI contributed by other galaxies during this time, (2) the largest instantaneous PI contributed by any individual galaxy during this time, (3) the total number of galaxies that contribute a PI greater than $\sim 10^{-5}$ (equivalent to a galaxy with a 1:10 mass ratio coming within $\sim 20~R_{\rm eff}$) during this time ($n_{\rm perturbers}$) and (4) the minimum separation from another galaxy with a mass ratio of at least 1:10 as a fraction of the galaxy effective radius. Reasonable changes to the size of the window (i.e. not so large as to wash out any signal or so small as to be unlikely to enclose the event responsible for the morphological disturbance) does not qualitatively change our conclusions.

\begin{figure*}
	\centering
    \includegraphics[width=0.95\textwidth]{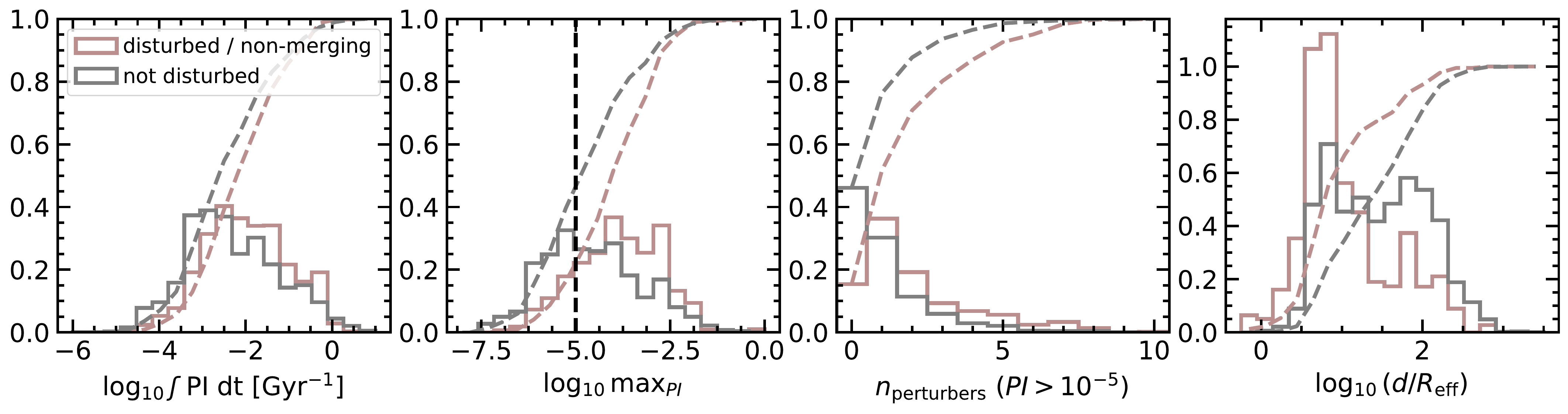}
    \caption{Histograms show the distribution of \textbf{Left}: the integrated perturbation index; \textbf{Centre left}: the largest instantaneous PI contributed by any individual galaxy; \textbf{Centre right}: the total number of galaxies that contribute a PI greater than $\sim 10^{-5}$; \textbf{Right}: the minimum separation from another galaxy with a mass ratio of at least 1:10 for the disturbed and mass-matched non-disturbed samples. Light red histograms indicate the distribution of quantities of the non-merging disturbed sample and grey histograms indicate the distribution of quantities of the non-disturbed sample. Dotted lines indicate the cumulative distribution.}
    \label{fig:PI_test}
\end{figure*}

The distribution of each of these quantities is shown in Fig \ref{fig:PI_test}. We find that the total integrated PI (left panel) over this time is comparable for the disturbed and non-disturbed samples, indicating that the integrated tidal field is not typically responsible for producing these disturbances (this may be a more important effect in a volume featuring more massive clusters). However we observe an offset in the distribution of the largest instantaneous PI contributed by an individual galaxy (centre left panel), indicating that a single object imparts more significant perturbations on average in the disturbed sample. If we consider the total number galaxies that contribute significantly to the PI (centre right panel) we see that only $\sim 15$ per cent of disturbed galaxies have undergone no interaction of this kind within the 200~Myr time frame compared with $\sim 50$ per cent of non-disturbed galaxies. Finally, if we consider the minimum separation from any object with at least a 1:10 mass ratio during the 200~Myr time frame (right panel), we find considerably closer separations on average for disturbed galaxies (peaking around $6~R_{\rm eff}$). Together these results indicate that, other than mergers, the main source of morphological disturbances are close encounters with individual galaxies (i.e. `fly-bys') rather than other processes like the integrated tidal contribution of galaxies in groups or clusters.

\section{Asymmetry and Gini/M$_{20}$ parameters}
\label{sec:merger_test}

In order to probe the observability of merging and disturbed dwarf galaxies over cosmic time, we consider two commonly used non-parametric methods for identifying merging galaxies, asymmetry \citep{Conselice2000} and $Gini$-$M_{20}$ \citep{Lotz2004}.

\subsection{Images}

We first produce mock observations in the rest-frame $r$-band as described in Section \ref{sec:mock_images} and \citet{Kaviraj2017}. Images are produced for every $R>1:10$ merger in 50~Myr increments between 100~Myr before $t_{\rm max}$ and 700~Gyr after $t_{\rm max}$. We also produce mock images for all galaxies at 8 snapshots between $z=0.5$ and $z=4$. For all disturbed galaxies at these snapshots we also produce mock images between 200~Myr and 200~Myr after in 50~Myr increments. All images are re-scaled to a fixed angular pixel scale of 0.05" (the approximate angular resolution of the JWST NIRCam instrument \citealt{Gardner2006}) as well as a fixed physical scale of 400~pc. The value of 400~pc is chosen so that the physical and angular pixel scales match at $z=1$. We do not attempt to add noise or account for exposure time, point spread function, seeing etc., since these will all be particular to the instrument used.

\subsection{Asymmetry}

To calculate the asymmetry, we follow a similar procedure to \citet{Conselice2000, Conselice2003} and calculate the shape asymmetry, $A_{g}$, using rest-frame $r$-band mock images. Mock images are smoothed by Gaussian kernel with a standard deviation equal to $\sfrac{1}{6}$ of the radius defined by the $\eta$ function for $R(\eta = 0.2)$ \citep[][]{Petrosian1976} and the shape asymmetry is calculated within 1.5~$R(\eta = 0.2)$ as follows:

\begin{equation}
    \label{eqn:asym}
    A_{g} = \frac{\sum_{i,j}|I(i,j)-R(i,j)|}{2\ \sum_{i,j}|I(i,j)|},
\end{equation}

\noindent Where $I$ is the smoothed image and $R$ is $I$ rotated by 180 degrees. The centre of rotation is chosen so as to minimise $A_{g}$ following \citet{Conselice2000}. We do not attempt to account for the effects of noise, surface brightness dimming and other observational biases on the selection of mergers using the shape asymmetry, which would likely act to reduce observability timescales at higher redshift.

Although it is not used in the merger classification, we also calculate the concentration index, $C$ \citep{Conselice2003}, again within 1.5~$R(\eta = 0.2)$:

\begin{equation}
    C = 5\ {\rm log_{10}}\left( \frac{r_{80}}{r_{20}} \right),
\end{equation}

\noindent where $r_{80}$ and $r_{20}$ are circular radii containing 80 and 20 per cent of the total intensity. We define the centre of the object so that the asymmetry is minimised in the same way as Eqn \eqref{eqn:asym}. A higher value of $C$ corresponds to a larger fraction of total intensity being found in the central region compared to the outskirts.

\subsection{$Gini$-$M_{20}$}

To calculate the $Gini$ and $M_{20}$ coefficients, we follow \citet{Lotz2004}. We first compute the $Gini$ coefficient as follows:

\begin{equation}
    Gini = \frac{1}{2\ \langle f \rangle n (n-1)}\sum^{n}_{i=1} \sum^{n}_{j=1} |f_{i} - f_{j}|,
\end{equation}

\noindent where $\langle f \rangle$ is mean flux over all pixel values $f_{i}$ and $n$ is the number of pixels in the galaxy segmentation map such that $Gini$ is equal to 0 if all pixels have a uniform flux and equal to 1 if a single pixel contains all the flux. We note that the $Gini$ coefficient is sensitive to the segmentation map as including more faint pixels in the outskirts of the galaxy will push the $Gini$ coefficient closer to 1 versus a more conservative segmentation map that includes only the brighter parts of a galaxy.

We then compute $M_{20}$, the second moment of the brightest 20 per cent of galaxy flux by first sorting pixels from highest to lowest intensity, we then compute $M_{20}$ as follows:

\begin{equation}
    M_{20} = {\rm log_{10}}\left( \frac{\sum_{i} M_{i}}{M_{tot}} \right)\ {\rm while}\ \sum_{i}f_{i}<0.2 f_{\rm tot},
\end{equation}

\noindent where $M_{i}$ is the intensity-weighted central second-moment of each pixel, $M_{tot}$ is the total intensity-weighted central second-moment, $f_{i}$ are the sorted intensities of each pixel and $f_{\rm tot}$ is the total intensity of the image. Whereas the $Gini$ coefficient indicates how evenly spread flux is throughout the object, $M_{20}$ indicates how bright pixels are distributed relative to the galaxy centre. 

\subsection{Mergers and disturbed galaxies in $A_{g}$-$C$ and $Gini$-$M_{20}$}


Here, we consider the position of mergers and disturbed galaxies in the $A_{g}$-$C$ and $Gini$-$M_{20}$ plane. We illustrate the point at which each merger is most clearly detectable by plotting the maximum values of $A_{g}$ and $Gini$ over the course of the whole merger rather than their instantaneous value at a given snapshot. Similarly, for disturbed galaxies, we take the maximum value between 200~Myr before and after a given snapshot. The maximum value of $A_{g}$ is typically reached at the same time the maximum value of $Gini$ (within 50~Myr of each other 64 per cent of the time), meaning we can safely assume that galaxies mergers are most detectable according to either method at roughly the same time. In order to improve statistics for the merging galaxy sample we include all mergers within $\pm50$~Myr of the snapshot in question.

In order to calibrate the observability timescales for $A_{g}$-$C$ and $Gini$-$M_{20}$ presented in Fig \ref{fig:timescales}, we adopt merger thresholds for each parameter. For $A_{g}$, we use a threshold of $A_{g}>0.35$, which is commonly used in the literature to classify mergers \citep[e.g.][]{Conselice2009}. Since $Gini$ is relatively sensitive to depth, resolution and scale of the image, we do not adopt the usual  threshold for mergers ($Gini>-0.14\ M_{20}+0.33$) \citep[e.g.][]{Lotz2004}, which is calibrated on galaxies in the low redshift Universe. We instead allow the normalisation of the relation between $Gini$ and $M_{20}$ to vary so that 80 per cent of merging galaxies lie above the threshold when the value of $Gini$ is at its maximum (as illustrated by the red dashed line in Fig \ref{fig:Gini_M20}).

\begin{figure}
	\centering
    \includegraphics[width=0.45\textwidth]{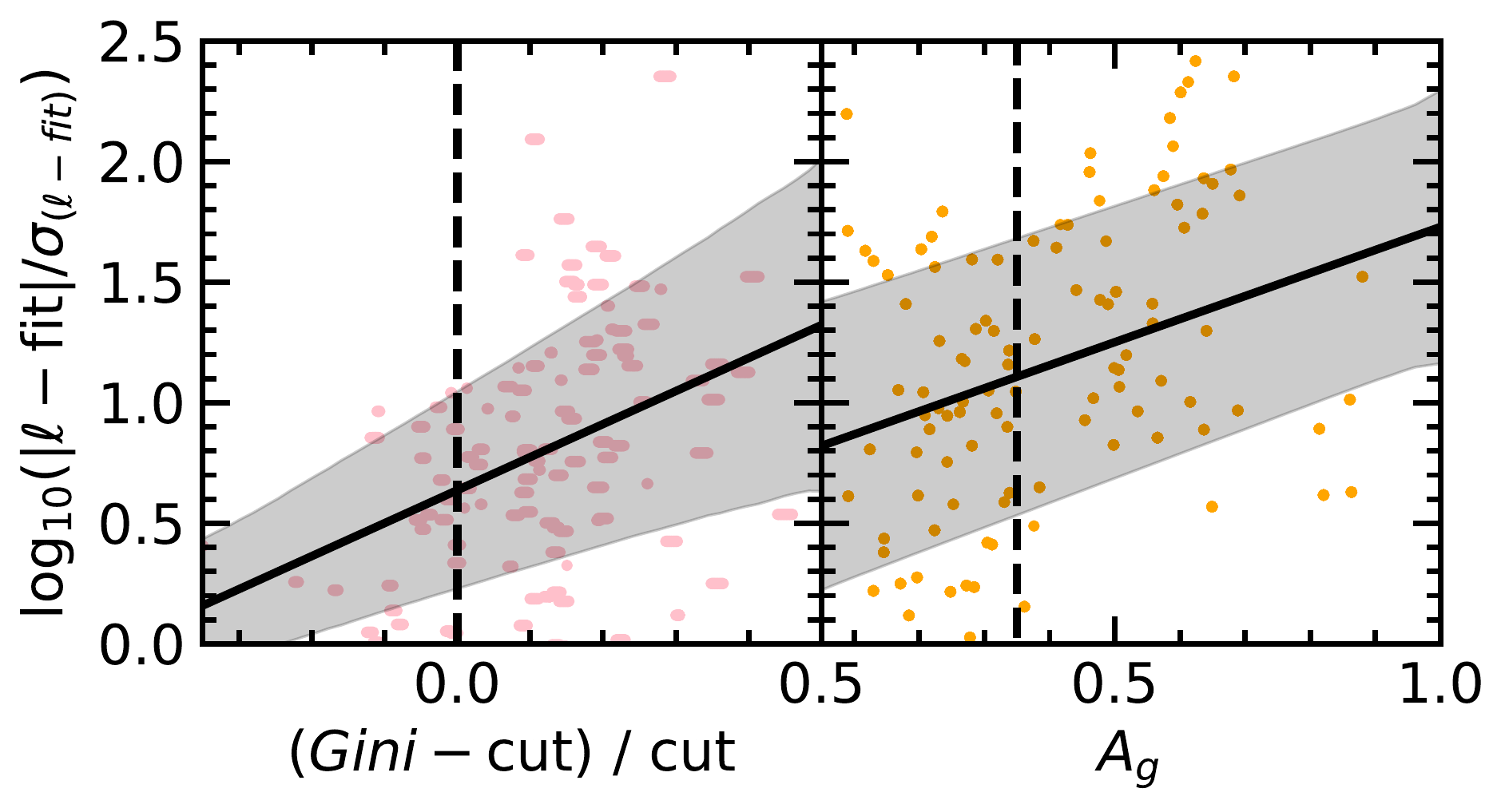}
    \caption{The residual between the smoothed continuum fit of the axis length and the actual evolution of the axis length as described in Section \ref{sec:disturbances_def} vs the distance above the $Gini$-$M_{20}$ cut (left) and shape asymmetry (right). The black lines show a first order polynomial fit to the data and the shaded region shows the standard deviation of the residual. The dashed lines indicate the threshold above which galaxies are considered to be merging.}
    \label{fig:compare_disturbances_to_residual}
\end{figure}

Fig \ref{fig:compare_disturbances_to_residual} shows the residual between the smoothed continuum fit of the axis length and the actual evolution of the axis length as described in Section \ref{sec:disturbances_def} vs the distance above the $Gini$-$M_{20}$ cut in the left panel and shape asymmetry in the right panel. In both cases, there is some correspondence between the value of the residual and the value of the non-parametric measures, albeit with considerable scatter in both measures. In the case of $A_{g}$ around 50 per cent of mergers that we consider morphologically disturbed would not be identified with an $A_{g}>0.35$ cut.

\begin{figure*}
	\centering
    \includegraphics[width=0.95\textwidth]{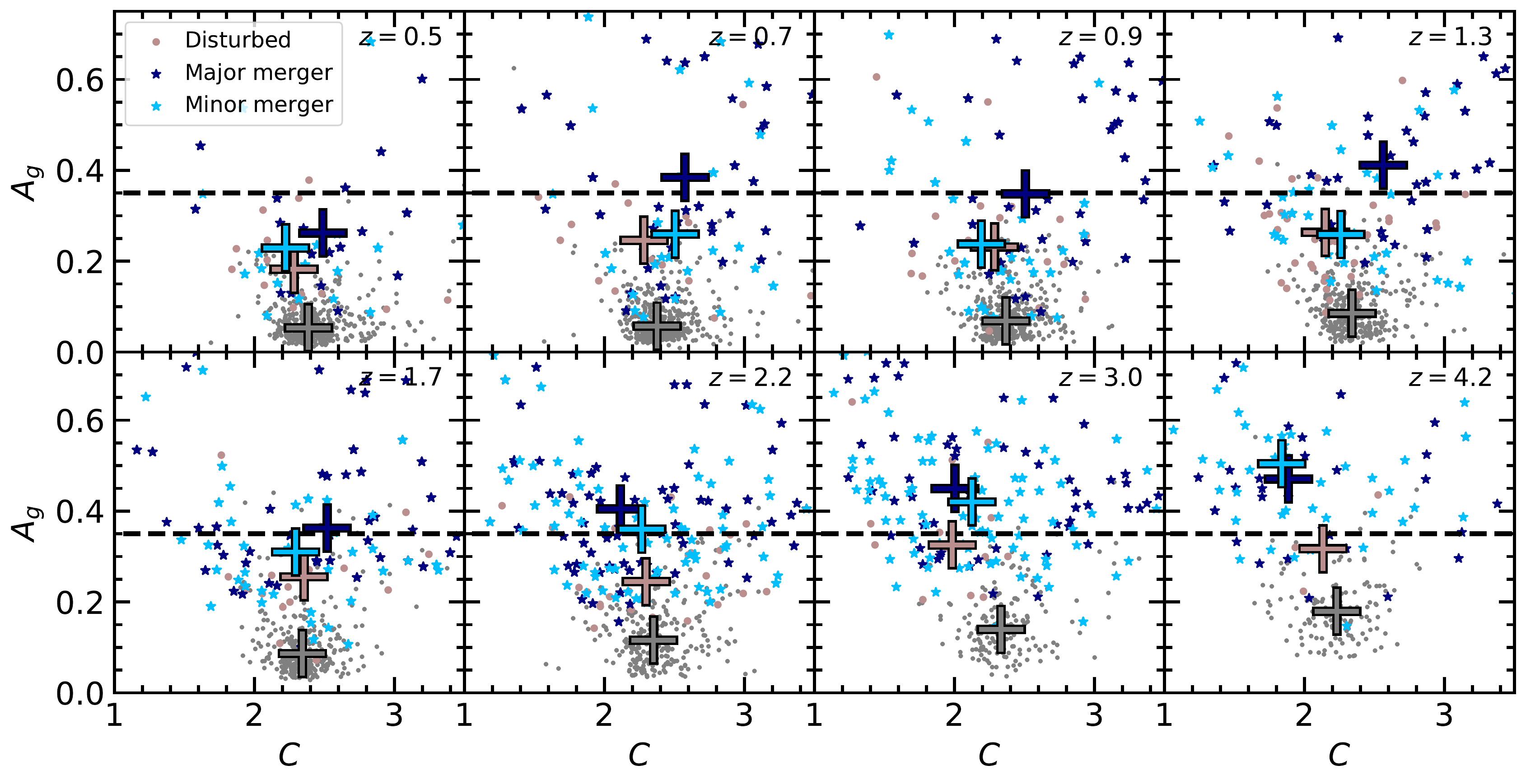}
    \caption{Shape asymmetry vs concentration for galaxies at different redshift snapshots. Dark blue and light blue stars indicate galaxies that undergoing either major or minor mergers, light red points indicate morphologically disturbed galaxies and grey points indicate galaxies that are neither merging or morphologically disturbed. Large coloured crosses indicate the mean position of each sample.}
    \label{fig:shape_A}
\end{figure*}

Figure \ref{fig:shape_A} shows the distribution of $A_{g}$ and $C$ for undisturbed dwarf galaxies (grey), dwarf galaxies that are undergoing major (dark blue) or minor (light blue) mergers and for disturbed dwarf galaxies (light red). While non-merging galaxies are not typically asymmetrical enough to be confused for mergers, around half of major mergers and the majority of minor mergers never become asymmetrical enough over the course of the merger to reach the threshold of $A_{g}>0.35$ indicated by the dashed black line. Disturbed galaxies are, on average somewhat more asymmetrical than non-merging galaxies but less asymmetrical than merging galaxies. There is significant crossover with minor mergers, however.

\begin{figure*}
	\centering
    \includegraphics[width=0.95\textwidth]{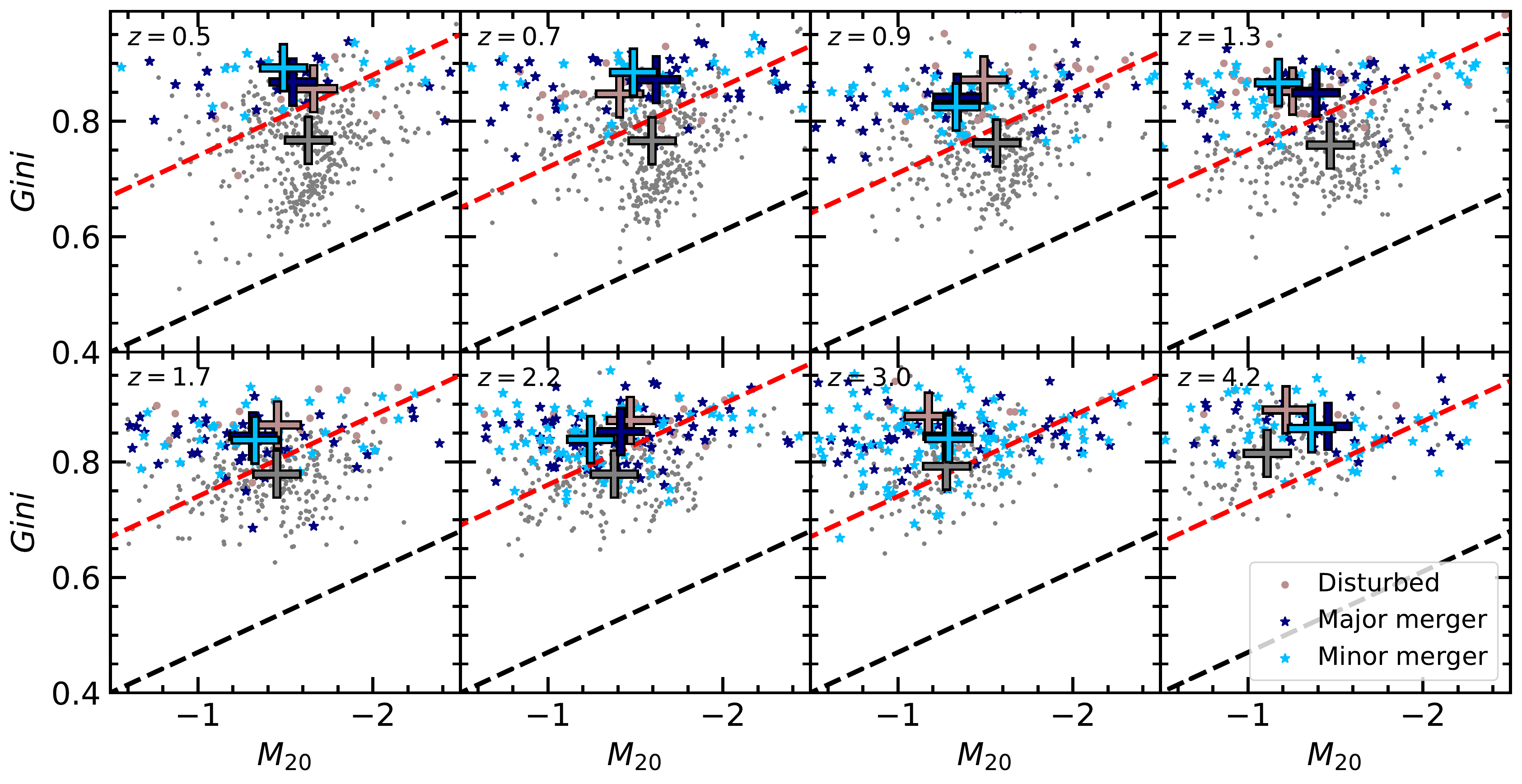}
    \caption{$Gini$ vs $M_{20}$ for dwarf galaxies at different redshift snapshots. Dark blue and light blue stars indicate galaxies that undergoing either major or minor mergers, light red points indicate morphologically disturbed galaxies and grey points indicate galaxies that are neither merging or morphologically disturbed. Large coloured crosses indicate the mean position of each sample. The black dashed line shows the cut $Gini > -0.14 M_{20} + 0.33$ which is calibrated to identify mergers in the low redshift Universe and the red dashed line shows the cut that recovers 80 per cent of merging galaxies.}
    \label{fig:Gini_M20}
\end{figure*}

Similarly, Fig \ref{fig:Gini_M20} shows shows the distribution of $Gini$ and $M_{20}$ for undisturbed dwarf galaxies (grey), dwarf galaxies that are undergoing major (dark blue) or minor (light blue) mergers and for disturbed dwarf galaxies (light red). non-merging dwarfs generally appear well separated from merging and disturbed galaxies, with major mergers, minor mergers and disturbed galaxies inhabiting the same area of the parameter space.

Although mergers and morphologically disturbed dwarfs can be relatively cleanly separated from non-merging galaxies using $Gini$-$M_{20}$ and $A_{g}$-$C$, it is still difficult to separate disturbed and merging galaxes using non-parametric measures of galaxy structure, particularly minor mergers, which inhabit a similar region of parameter space to morphologically disturbed dwarfs.

\section{Testing the dependence of our results on choice of parameters and timescales}
\label{sec:test}

In this section we briefly explore how our choice of the parameters used to determine if objects are morphological disturbed (see Section \ref{sec:disturbances_def}) affect merger durations. We also explore how changes in the merger duration used to calculate the fraction of merger driven star formation in Eqn \eqref{eqn:xi}, affects our results.

\begin{figure}
	\centering
    \includegraphics[width=0.45\textwidth]{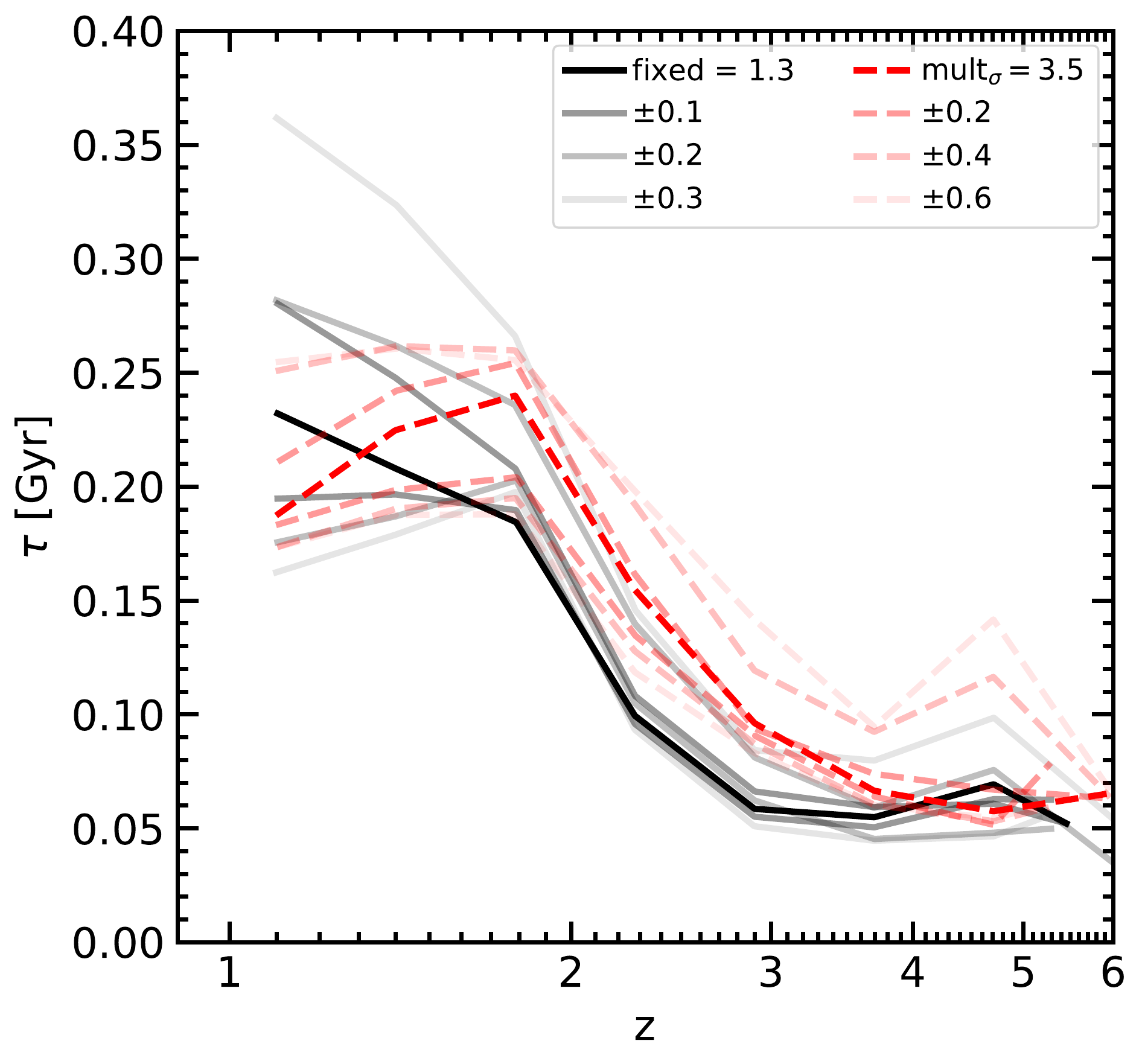}
    \caption{Evolution of median major merger durations as a function of redshift for different input parameters. Red dashed lines indicate merger timescales when we use a threshold based the multiple of $\sigma_{\rm local}$ that axis lengths are allowed to vary from the continuum fit before they are considered morphologically disturbed. Black solid lines indicate the same but using a fixed threshold rather than one that varies with $\sigma_{\rm local}$. Lighter lines show how merger durations vary for small changes in the threshold.}
    \label{fig:timescale_fixed}
\end{figure}

Fig \ref{fig:timescale_fixed} shows how merger durations vary when we consider a threshold based on mult$_{\sigma}$, which is the multiple of $\sigma_{\rm local}$ that axis lengths are allowed to vary from the continuum fit before they are considered morphologically disturbed (as used in Section \ref{sec:disturbances_def}) or using a fixed threshold that does not vary with $\sigma_{\rm local}$. These are indicated by red dashed and black solid lines respectively. At lower redshifts ($z<3$), we can pick values of either threshold that produce essentially identical evolution in the merger duration as a function of redshift (different thresholds are indicated as lines that become lighter as the threshold moves further from our adopted value). However, there is a small depression in merger durations at higher redshifts when we used a fixed threshold. Increasing or decreasing either threshold by a small amount does not qualitatively alter the trend with redshift and only slightly alters the normalisation.

\begin{figure}
	\centering
    \includegraphics[width=0.45\textwidth]{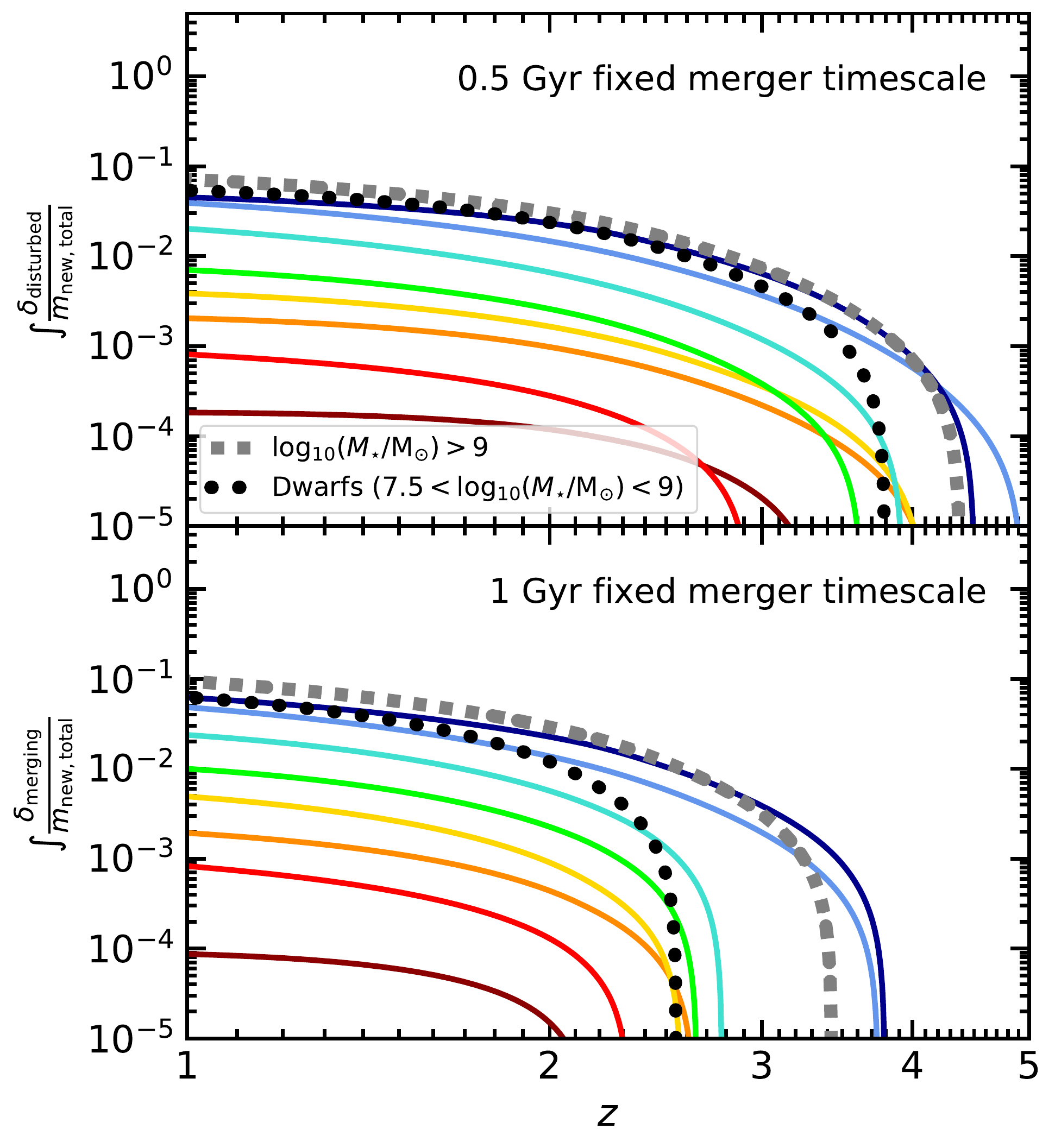}
    \caption{Version of Fig \ref{fig:budget} using a fixed merger timescale of 0.5 Gyrs (top panel) and 1 Gyr (bottom panel).}
    \label{fig:budget_fixed}
\end{figure}

Fig \ref{fig:budget_fixed} shows the cumulative star formation budget due to mergers when we use 0.5~Gyr and 1~Gyr fixed timescales (the same as the fourth panel of Fig \ref{fig:budget}). In principle, since the star formation excess is calculated relative to a non-merging sample it should be relatively robust to the exact choice of timescale, since an increase in timescale should decrease the excess relative to the non-merging population but also increase the total stellar mass formed in the merging sample and vice-versa. Although the exact evolution differs, we find that the final $z=1$ budgets are indeed relatively unaffected by our choice of merger duration. We find that 6 per cent / 9 per cent of stellar mass is driven by mergers in massive galaxies and dwarfs respectively using a fixed merger duration of 0.5~Gyrs. For a fixed duration of 1~Gyr the fractions are almost identical (6 / 10 per cent). For the dwarf and $M_{\star}>10^{9}$M$_{\odot}$ samples, the results do not differ significantly from those in Section \ref{sec:induced_star_formation}. The exact choice of timescale, therefore, does not qualitatively alter our conclusions. 








\bsp	
\label{lastpage}
\end{document}